\documentclass[aps,prd,twocolumn,superscriptaddress,amsmath,amssymb,longbibliography]{revtex4-2}

\usepackage{url}
\usepackage{graphicx}
\usepackage{xcolor}
\usepackage[normalem]{ulem}

\DeclareFontFamily{OMS}{oasy}{\skewchar\font48 }
\DeclareFontShape{OMS}{oasy}{m}{n}{%
         <-5.5> oasy5     <5.5-6.5> oasy6
      <6.5-7.5> oasy7     <7.5-8.5> oasy8
      <8.5-9.5> oasy9     <9.5->  oasy10
      }{}
\DeclareFontShape{OMS}{oasy}{b}{n}{%
       <-6> oabsy5
      <6-8> oabsy7
      <8->  oabsy10
      }{}
\DeclareSymbolFont{oasy}{OMS}{oasy}{m}{n}
\SetSymbolFont{oasy}{bold}{OMS}{oasy}{b}{n}

\DeclareMathSymbol{\smallleftarrow}     {\mathrel}{oasy}{"20}
\DeclareMathSymbol{\smallrightarrow}    {\mathrel}{oasy}{"21}
\DeclareMathSymbol{\smallleftrightarrow}{\mathrel}{oasy}{"24}

\begin{document}

\title{Neutrino flavor mixing with moments}

\author{McKenzie Myers}
\affiliation{Department of Physics, North Carolina State University, Raleigh, NC 27695, USA}

\author{Theo Cooper}
\affiliation{Department of Physics, North Carolina State University, Raleigh, NC 27695, USA}

\author{MacKenzie Warren}
\affiliation{AI Research Group, 5x5 Technologies, St Petersburg, FL 33701, USA}

\author{Jim Kneller}
\email{jim\_kneller@ncsu.edu}
\affiliation{Department of Physics, North Carolina State University, Raleigh, NC 27695, USA}

\author{Gail McLaughlin}
\affiliation{Department of Physics, North Carolina State University, Raleigh, NC 27695, USA}

\author{Sherwood Richers}
\altaffiliation{NSF Astronomy \& Astrophysics Postdoctoral Fellow}
\affiliation{Department of Physics, University of California Berkeley, Berkeley, CA 94720}

\author{Evan Grohs}
\affiliation{Department of Physics, North Carolina State University, Raleigh, NC 27695, USA}

\author{Carla Fr\"{o}hlich}
\affiliation{Department of Physics, North Carolina State University, Raleigh, NC 27695, USA}

\date{\today}

\begin{abstract}
The successful transition from core-collapse supernova simulations using classical neutrino transport to simulations using quantum neutrino transport will require the development of methods for calculating neutrino flavor transformations that mitigate the computational expense. One potential approach is the use of angular moments of the neutrino field, which has the added appeal that there already exist simulation codes which make use of moments for classical neutrino transport. Evolution equations for quantum moments based on the quantum kinetic equations can be straightforwardly generalized from the evolution of classical moments based on the Boltzmann equation. 
We present an efficient implementation of neutrino transformation using quantum angular moments in the free streaming, spherically symmetric bulb model. We compare the results against analytic solutions and the results from more exact multi-angle neutrino flavor evolution calculations. We find that our moment-based methods employing scalar closures predict, with good accuracy, the onset of collective flavor transformations seen in the multi-angle results. However in some situations they overestimate the coherence of neutrinos traveling along different trajectories. More sophisticated quantum closures may improve the agreement between the inexpensive moment-based methods and the multi-angle approach.
\end{abstract}

\maketitle

\section{Introduction}

The importance of neutrinos to the core-collapse supernova (CCSN) paradigm has been recognized since the earliest simulations of the explosions by Colgate and White \cite{1966ApJ...143..626C}. Even though many details have changed over the years since that pioneering study, neutrinos are still thought to be the driver of the explosions of stars with initial masses $\gtrsim 10\,M_\odot$ that have reached the end of their nuclear burning lifetimes (see \cite{bethe_supernova_1990,kotake_explosion_2006,janka_theory_2007,janka_explosion_2012,burrows_colloquium_2013,foglizzo_explosion_2015} for reviews). 

Although a complete theoretical accounting of the explosion dynamics remains elusive, state of the art numerical simulations performed in three dimensions appear to be converging toward exploding solutions \cite{lentz_three-dimensional_2015,muller_supernova_2017,ott_progenitor_2018,summa_rotation-supported_2018,oconnor_exploring_2018,nakamura_long-term_2019,glas_effects_2019,glas_three-dimensional_2019,burrows_overarching_2020}. However, there are still many uncertainties underlying these models. The structure of the progenitor stars is presently poorly understood and is complicated by variations in stellar mass, metallicity, rotation history, magnetic fields, and binary interactions. Another source of uncertainty in the details of the explosion arises from the unknown equation of state describing matter above nuclear densities. Variations in the equation of state have been shown to lead to significant differences in the outcomes of simulations (e.g., \cite{2020PhRvL.124i2701Y,2020ApJ...894....4D,2021arXiv210713016G}). Furthermore, there is yet work to be done to ensure that multidimensional simulations are performed with sufficient resolution and that the approximations employed in various methods (especially regarding neutrino transport) do not significantly affect the solution \cite{radice_neutrino-driven_2016,richers_detailed_2017}, though there currently seems to be more agreement than disagreement between full supernova simulation codes \cite{oconnor_global_2018,just_core-collapse_2018}.

While the recent progress in supernova simulations is impressive, the simulations neglect the quantum nature of the neutrino and its ability to change flavor (though see \cite{dasgupta_role_2012,stapleford_coupling_2020} for some attempts at effective treatments). Until recently, this omission was not thought to be important for the dynamics of the explosion (see \cite{duan_collective_2010,bellini_neutrino_2014} for recent reviews). Flavor transformation calculations based on post-processing results from simulations that employ classical transport reveal a number of neutrino flavor transformation phenomena. As in the Sun, neutrinos undergo complete flavor transformation at a Mikheyev-Smirnov-Wolfenstein (MSW) resonance \cite{Mikheyev:1985aa,Mikheyev:1986tj,1978PhRvD..17.2369W}  due to the combination of differing neutrino masses and a potential from neutrinos interacting with the background matter. However, this flavor transformation occurs well outside of any supernova engine or nucleosynthesis region and is mostly important for understanding future neutrino detections at Earth, e.g. \cite{Kneller:2007kg,Gava:2009pj,Patton:2013dba}. The neutrino self-interaction potential due to neutrinos interacting with other nearby neutrinos makes the flavor transformation nonlinear. The so-called collective neutrino oscillations, which result from the combined effects of the self-interaction potential and the differing neutrino masses, are also thought to occur too far out to affect the supernova engine in a significant way \cite{chakraborty_no_2011,duan_self-induced_2011,dasgupta_role_2012}, unless beyond the standard model interactions are included, e.g. \cite{Stapleford:2016jgz}. 

But more recently, the so-called fast flavor instability was shown to have the potential to drive significant and rapid flavor transformations well inside the shock if the neutrino distribution fulfils the criterion of an electron lepton number crossing \cite{sawyer_speed-up_2005,sawyer_neutrino_2016,chakraborty_self-induced_2016,izaguirre_fast_2017,capozzi_fast_2017,dasgupta_fast_2018,abbar_fast_2018,Johns:2019izj,Capozzi:2020syn,Johns:2021taz,Nagakura:2021hyb,Nagakura:2021txn,Nagakura:2021hyb}. Conditions fulfilling this instability criterion have been shown to exist both inside \cite{abbar_fast_2019,shalgar_occurrence_2019,m_delfan_azari_fast_2020,glas_fast_2020} and outside \cite{morinaga_fast_2020} the shock front in simulations that use classical transport. Recent consideration of effects of small-scale turbulence could make the fast flavor instability even more ubiquitous than the electron lepton number crossing criterion predicts \cite{Mangano:2014zda,Abbar:2020ror,Sigl:2021tmj}. Thus these more recent studies now strongly indicate that flavor transformation may occur in regions of the supernova where there could be feedback into the hydrodynamics. While simulations that adopt effective treatments of flavor transformation within the framework of classical neutrino transport may be able to capture such physics in an approximate way, direct calculations of quantum neutrino transport are needed to understand the effects of microscopic instabilities from first principles. 

There are several challenges to be overcome in implementing and simulating quantum neutrino transport. One unavoidable challenge is a huge increase in the dynamical range of lengthscales that need to be resolved. At the present time, the spatial resolution of the state-of-the-art simulations approaches lengthscales of order $\sim~100\;{\rm m}$ for a simulation covering a domain size of order $\sim 1000\;{\rm km}$.
However, the lengthscale of neutrino flavor oscillations in matter with a density of $\sim 10^{12}\;{\rm g/cm^3}$ is of order $\sim 1\;{\rm \mu m}$, which means one would need a simulation with a dynamical range of at least twelve orders of magnitude. 
Multi-angle collective flavor transformation calculations also require hundreds to thousands of angle bins in order for the results to converge, far more than the $\sim 10$ angle bins often used in simulations of classical transport \cite{2008CS&D....1a5007D,2012PhRvD..86l5020S,richers_detailed_2017,iwakami_simulations_2020}. 
The results from multi-angle calculations are often seen to exhibit substantial changes seen as the number of angle bins used is changed. 
The number of energy bins used in multi-angle calculations is also usually of order a few hundred, an order of magnitude larger than the $\sim 20$ energy bins typically used in classical simulations. However, in contrast to the number of angle bins, the convergence of the multi-angle results with the the number of energy bins is often observed to be much smoother \cite{2012PhRvD..86l5020S} and the large number of energy bins used is driven by the desire for sufficient resolution of the spectrum. Thus the computational expense of even a 1D, spherically symmetric supernova simulation using quantum transport is expected to be many orders of magnitude greater than for a simulation using classical transport. One must seriously consider alternative approaches that mitigate this expense if quantum supernova simulations are to be feasible. 

One approach, which we consider here, is to use angular moments of the quantum neutrino distribution i.e. quantum moments. 
{There are good motivations for considering this approach. First, the coupling of the neutrinos to the rest of the fluid is most simply expressed via the moments so computing the moments directly is more efficient than computing the flavor evolution of neutrinos traveling along different trajectories and then integrating.} 
Second, neutrino transport based on angular moments of the classical neutrino radiation field are already used in many state of the art supernova simulation codes due to their computational efficiency (e.g ., \cite{jiang_algorithm_2014,cimerman_hydrodynamics_2017,birindelli_high_2017,hopkins_numerical_2019,bloch_high-performance_2021,fuksman_two-moment_2021}). 
Modifying such codes to include the equations describing the evolution of the quantum moments will rely on techniques developed both in the neutrino oscillation literature and in the neutrino transport literature. \citet{vlasenko_neutrino_2014} and \citet{volpe_neutrino_2015} developed the equations from first principles, while \citet{blaschke_neutrino_2016} expressed the collision integral in detail. Investigations of the feasibility of the quantum moment approach have already been taken. \citet{strack_generalized_2005} (see also \citet{2014JCAP...10..084D}) outlined a moment-based formalism for the QKEs and \citet{richers_neutrino_2019} fleshed out the form of the collision integral using interaction rates in the form commonly used in the core-collapse supernova simulation literature. \citet{richers_neutrino_2019} also developed a code to simulate the QKEs under the assumption of isotropy and homogeneity, and \citet{johns_fast_2020} used a moment method to analyze the presence of fast flavor instability in parameterized cases. 
When comparing moment-based approaches to more-sophisticated transport, disagreement emerges from the need to truncate the tower of moment evolution equations at some level yielding one or two fewer equations than the number of unknown moments.
To solve the equations one must propose an algorithm to estimate the unknown moment(s) given the evolved moments. If this algorithm were perfect, the evolved moments are guaranteed to exactly match those extracted from a full Boltzmann calculation. Truncating the tower of moment equations at different levels does not necessarily lead to more accurate results. The accuracy of the results 
is determined by the accuracy of the algorithm for finding the un-evolved moment(s).

The goal of this paper is to present results from a new code that solves the quantum moment evolution equations for a supernova neutrino bulb model and allows us to explore these issues. In Section~\ref{sec:flavor_moment_scheme} we present the quantum kinetic moment equations we use,  introduce the one-moment and two-moment schemes and explain the closures we adopt for them. In Section~\ref{sec:tests} we investigate how well moment methods are able to produce MSW and collective oscillations by comparing their results to  multi-angle calculations. Finally, we conclude in Section~\ref{sec:conclusions} and discuss the issues our study uncovers. 

\section{Flavor mixing with moments}
\label{sec:flavor_moment_scheme}

Classical radiation transport is described in general by the Boltzmann equation, but the six-dimensional distribution function is exceedingly computationally expensive to evolve. A moment approach to radiation transport involves taking angular moments of the full Boltzmann equation and evolving a subset of these moments rather than directly discretizing the neutrino distribution in angle. The result is a system of coupled differential equations for a handful of four-dimensional (three spatial and one energy dimension) variables, rather than a differential equation for a six-dimensional distribution function. The angular moments which are evolved or prescribed in a two-moment method in flat spacetime are the differential energy density $E$, (energy) flux vector $\vec{F}$, and pressure tensor $\tensor{P}$, defined as 
\begin{align}
E(t,\vec{r},q) &= \frac{1}{4\pi}\,\left(\frac{q}{2 \pi \hbar c}\right)^{3} \int d\Omega_{p}\,f(t,\vec{r},\vec{p}) \\
\vec{F}(t,\vec{r},q) & = \frac{1}{4\pi}\,\left(\frac{q}{2 \pi \hbar c}\right)^{3}  \int d\Omega_{p}\,\hat{p}\, f(t,\vec{r},\vec{p})\\
\tensor{P}(t,\vec{r},q) & = \frac{1}{4\pi}\,\left(\frac{q}{2 \pi \hbar c}\right)^{3}  \int d\Omega_{p}\, \hat{p}\otimes \hat{p}\, f(t,\vec{r},\vec{p})\,\,,
\end{align}
where $q = |\vec{p}|\,c$. Note that we omit the speed of light from the definition of the flux moment in order that the moments all have the same units. The factor of $1/4 \pi$ in each definition of the moments is a units choice made to divide out the $4\pi$ from the solid angle integral when tracking the evolution of these variables. This means, the total neutrino flux $\vec{\cal{F}}$ is $4 \pi$ times the integral of the spectral flux 
\begin{equation}
    {\vec{\cal{F}}} = 4\pi\, \int_0^\infty {\vec{F}}\,dq\,\,.
\end{equation}
The moments for the antineutrinos are defined similarly and shall be denoted with overbars i.e. $\bar{E}$, $\vec{\bar{F}}$ and $\tensor{\bar{P}}$. 
In classical transport $f$ and $\bar{f}$ are distribution functions, taking on values $f\in[0,1]$ and describing the occupation number per unit of phase space. To generalize so as to permit flavor mixing, we refine $f$ and $\bar{f}$ to become matrices in flavor space. Thus, the moments also become matrices and the energy density, for example, is now (assuming two neutrino flavors) \cite{strack_generalized_2005}
\begin{equation}
    E(t,\vec{r},q) \rightarrow \begin{pmatrix} 
    E^{(ee)}(t,\vec{r},q) & E^{(ex)}(t,\vec{r},q) \\
    E^{(xe)}(t,\vec{r},q) & E^{(xx)}(t,\vec{r},q)
    \end{pmatrix}\,\,.
    \label{eq:energy-matrix}
\end{equation} 
The other moments have a similar structure. We shall refer to moments that are matrices in flavor space as `quantum moments' in order to distinguish them from the scalar `classical' definition of moments, and indicate flavor elements of the quantum moments with superscripts in parentheses to avoid confusions with exponents. $E^{(ee)}$ and $E^{(xx)}$ are real, positive definite quantities that represent the differential energy densities in the electron and representative heavy lepton (\lq\lq $x$") flavor neutrinos. The off-diagonal components $E^{(ex)}$ and $E^{(xe)}$ represent the flavor overlap and are complex quantities. The luminosity matrices of neutrinos and antineutrinos are calculated from the radial component $F_r$ and $\bar{F}_r$ of the flux vectors for the neutrinos and antineutrinos respectively, and are 
\begin{align}
L & = \left( 4 \pi r \right)^2 c \int F_r\,dq, \label{eq:luminositymatrix} \\
\bar{L} & = \left( 4 \pi r \right)^2 c \int \bar{F}_r\,dq\,\,.
\label{eq:luminositymatrixbar}
\end{align}
Similarly to the quantum moments for the energy density, flux and pressure, the diagonal elements of the luminosity matrices are the luminosities of each neutrino flavor. 

The evolution equations for the moments are found by following the procedure outlined in \citet{strack_generalized_2005} and \citet{zhang_transport_2013} (although the collision terms therein need to be modified as in \cite{blaschke_neutrino_2016,richers_neutrino_2019}). There are an infinite number of angular moments and corresponding evolution equations (e.g. \cite{thorne_relativistic_1981}). Under the assumption of flat spacetime and spherical symmetry, the evolution equations for the lowest-order moments (energy density and energy flux) take the form 
\begin{widetext}
\begin{align}
\frac{1}{c}\frac{\partial E}{\partial t} + \frac{\partial F_{r}}{\partial r} + \frac{2 F_{r}}{r}  &= -\frac{i}{\hbar\,c} \left[ H_{V}+H_{M} + H_E,E \right]-\frac{i}{\hbar\,c} \left[ H_F,F_{r}\right] +C_{E}\label{eq:energy} \\
\frac{1}{c}\frac{\partial \bar{E}}{\partial t} + \frac{\partial \bar{F}_{r}}{\partial r} + \frac{2 \bar{F}_{r}}{r}  &= -\frac{i}{\hbar\,c} \left[ H_{V}-H_{M}-H_{E}^{*},\bar{E} \right] +\frac{i}{\hbar\, c} \left[ H_{F}^{*},\bar{F}_{r}\right] +\bar{C}_{E}\label{eq:energybar} \\
\frac{1}{c}\,\frac{\partial F_{r}}{\partial t} + \frac{\partial P_{rr}}{\partial r} + \frac{3\,P_{rr} - E}{r}  &= - \frac{i}{\hbar\,c} \left[H_{V}+H_{M}+H_{E},F_{r}\right] -\frac{i}{\hbar\,c} \left[H_{F},P_{rr}\right] + C_{F}\label{eq:flux} \\
\frac{1}{c}\,\frac{\partial \bar{F}_{r}}{\partial t} + \frac{\partial \bar{P}_{rr}}{\partial r} + \frac{3\,\bar{P}_{rr} -\bar{E}}{r}  &= - \frac{i}{\hbar\,c} \left[H_{V}-H_{M}-H_{E}^{*},\bar{F}_{r}\right] +\frac{i}{\hbar\,c}  \left[H_{F}^{*},\bar{P}_{rr}\right] + \bar{C}_{F}\label{eq:fluxbar}
\end{align}
\end{widetext}
where we have defined
\begin{equation}
H_{E} =\phantom{-} 4 \pi\,\sqrt{2}\,G_{F}\,\int \frac{dq}{q} \left ( E(q) - \bar{E}^{*}(q) \right)
\label{eq:HSI_E}
\end{equation}
\begin{equation}
H_{F} = -4 \pi\,\sqrt{2}\,G_{F}\,\int \frac{dq}{q} \left ( F_r(q) - \bar{F}_r^{*}(q) \right)
\label{eq:HSI_F}
\end{equation}
and $^{*}$ indicates the complex conjugate. Together $H_E$ and $H_F$ are the self-interaction Hamiltonian $H_{SI} = H_E + H_F$. The four terms $C_{E}$, $\bar{C}_{E}$, $C_{F}$ and $\bar{C}_{F}$ are the `collision' terms and, in the context of flavor mixing, are also matrices \cite{blaschke_neutrino_2016,richers_neutrino_2019}. We neglect the collision terms in the rest of this work. For two flavor mixing, the vacuum Hamiltonian has the form
\begin{equation}
H_{V} = \frac{\Delta m^{2}_{12}\,c^4}{4\,q}\,\left[\, \sin(2\,\theta_{12})\,\sigma_{1}  -  \cos(2\,\theta_{12})\,\sigma_{3}\, \right]\,\,,
\label{eq:H_0}
\end{equation}
where $\Delta m_{12}^2=m_2^2 - m_1^2$ is the splitting between the neutrino masses and $\theta_{12}$ is the neutrino vacuum mixing angle. The matter Hamiltonian has the form
\begin{equation}
H_{M} = \frac{\sqrt{2} G_{F} n_{e}}{2} \sigma_{3}~,
\label{eq:H_e}
\end{equation}
where $n_e$ is the number density of electrons. In both cases, $\sigma_{i}$ is the $i$'th Pauli matrix and we have removed terms that contribute only to the trace of the Hamiltonian, as they do not affect the flavor evolution.  

In order to solve for the evolution of the moments we need to truncate the tower of equations. In classical moment transport this is typically done at the first or second level i.e. Eqns. \ref{eq:energy} and \ref{eq:energybar}, or \ref{eq:energy} to \ref{eq:fluxbar}. But this truncation introduces a hurdle: inspection of the truncated tower of equations reveals that they contain, in general, one more moment than the number of equations allows us to solve for. Thus to solve the truncated tower we need to close the evolution equations by divining a relationship between the moments that are evolved and the moments which are not. 
A perspicacious choice for the closure will result in the agreement of moments with higher resolution methods. Conversely, choosing an approximate closure will only yield approximate solutions.
Furthermore, truncating the tower of equations at a higher level does not necessarily lead to more accurate solutions for the moments that are solved, unless the structure of the higher moments is known better than that of the lower moments. Rather, evolving more moments simply provides more information with which to construct  closures, though increased accuracy of the solutions is not guaranteed just because we use a closure with more inputs.

\subsection{Neutrino Bulb Model}

In order to test the ability of moment-based schemes to reproduce the flavor transformation from more sophisticated transport calculations, we need to compare results from the two approaches. Unfortunately there are not many test problems which allow such a comparison. 
The test cases we consider in this work focus upon the inhomogeneous environment around a spherical source of neutrinos i.e. a neutrino bulb.
Since this environment has been studied extensively due to its similarity to core-collapse supernovae, we can take advantage of results from other codes which have been developed for this scenario. In what follows we compare our results with the steady-state `multi-angle' BULB model calculations using the SQA code, which is briefly described in Appendix~\ref{sec:SQA}.

For the moment-based calculations we shall consider truncations at the first and second level of the equation tower i.e. a scheme where we solve for $E$ and $\bar{E}$, or $E$, $\bar{E}$, $F_r$ and ${\bar{F}}_r$. For each we need a closure.
In previous studies of classical moment transport, closures have been supplied from analytic physical approximations or ad-hoc prescriptions (see \cite{smit_closure_2000,murchikova_analytic_2017} for summaries) or characteristic methods (e.g.,  \cite{hayes_beyond_2003,hubeny_new_2007}). There is a wealth of physics buried in the choice of closure and any results will depend on this choice. We leave exploration of the sensitivity to the closure to future work and use here two geometrically motivated examples. 

\begin{figure}[b]
\centering
\includegraphics[width = 0.5 \textwidth]{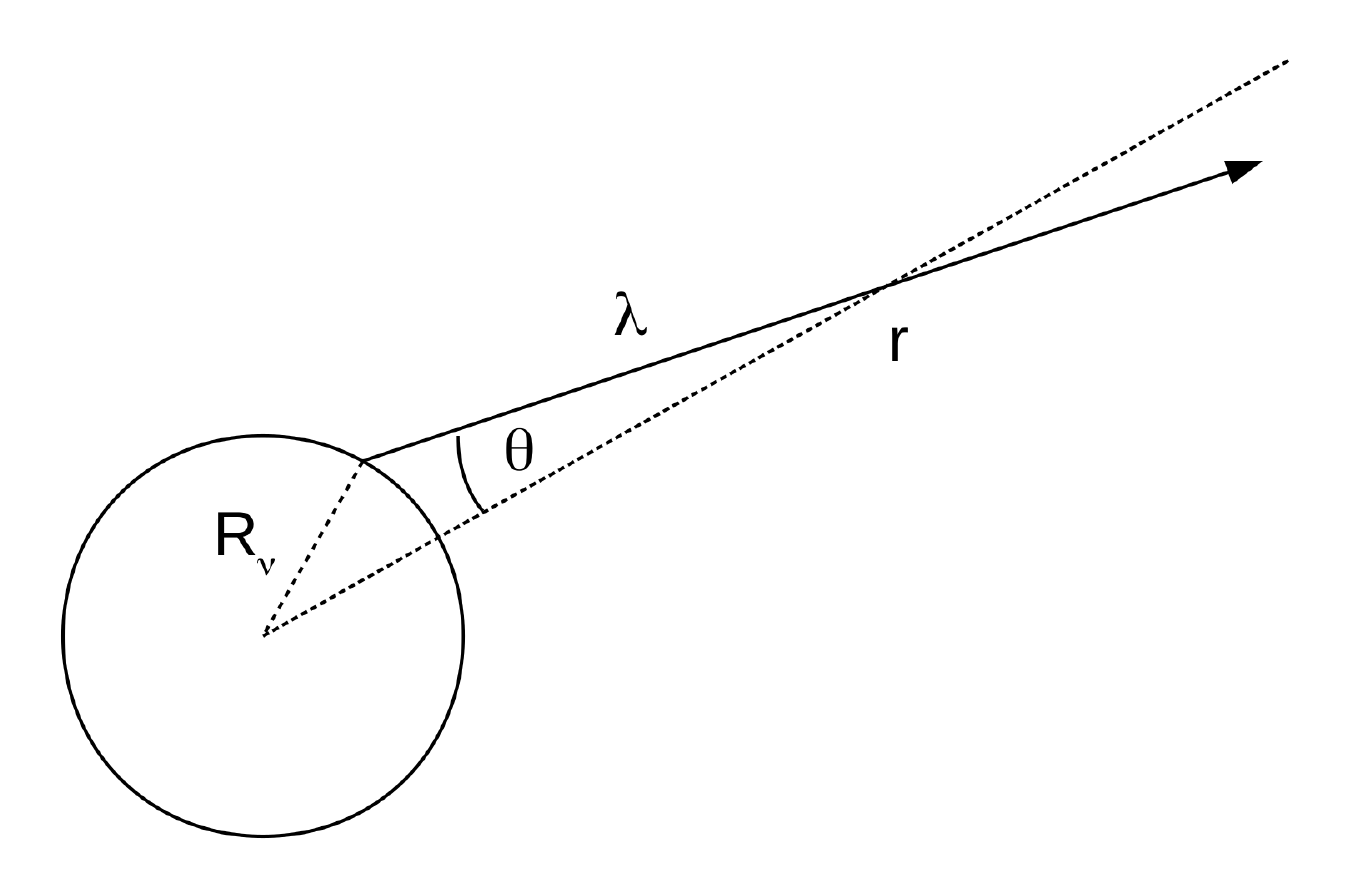}
\caption{The bulb model: at a distance $r$ from the center of a opaque neutrino bulb of radius $R_{\nu}$, the neutrino has traveled a distance $\lambda$ and its trajectory makes an angle $\theta$ with the radial direction.}
\label{fig:bulbmodel}
\end{figure}

The geometry of the scenario we are considering is shown in Figure~\ref{fig:bulbmodel}. At a given radial location $r$ above the source, the neutrinos are confined to propagate within a cone around the radial direction. Neutrinos which are propagating along a trajectory which makes an angle $\theta$ with the radial direction have traveled a distance 
\begin{equation}
\lambda = r\cos\theta - \sqrt{R_{\nu}^2 - r^2\,\sin^2 \theta}
\end{equation}
from the neutrinosphere $R_{\nu}$. The half opening-angle of the cone containing all the neutrino trajectories passing through the point of interest is $\theta_\mathrm{max}$ and from the geometry this angle is found to be
\begin{equation}
\cos \theta_\mathrm{max} = \sqrt{1 - (R_{\nu}/r)^{2}}. 
\label{eq:costhetamax}
\end{equation}
We will use this to self-consistently inform our choice of closures. 

\subsubsection{One-moment closure~\label{sec:single-angle}}
The first closure is a relationship between the energy and the radial component of the flux for use in one moment calculations. In classical flux-limited diffusion, the system is similarly closed so as to directly evolve only one moment, though the specific closure is different from that employed here. When the neutrinos are emitted half-isotropically from a spherical bulb and do not oscillate, the integrals defining the scalar energy density and radial flux (or more generally, the traces thereof) are related analytically. In this case we find the flux is related to the energy density via
\begin{equation}
    F_{r} = \frac{(1-\cos^{2}\theta_\mathrm{max})}{2\,(1-\cos \theta_\mathrm{max})} E~.
    \label{eq:1moment_closure}
\end{equation}
Using Equation~\ref{eq:costhetamax}, this can be used to define a closure as 
\begin{equation}
    F_r= \frac{ E}{2} \left(1+\sqrt{1 - \left(\frac{R_{\nu}}{r}\right)^{2}} \right).
    \label{eq:dighe}
\end{equation}
In what follows we will refer to this relationship as the ``one-moment'' closure. Substituting the one-moment closure of Eq.~\ref{eq:dighe} into Eq.~\ref{eq:energy} gives an equation that is now only a function of the two energy densities. Using this closure, the form of the self-interaction Hamiltonian matches that in the single-angle approximation of \citet{2008PhRvD..78c3014D}, although we stress that the advection terms in the one-moment approximation differ from those in the single-angle approximation.

\subsubsection{Two moment closure}

In the two moment approximation we simultaneously solve for the evolution of the first two moments as is done in M1 neutrino transport methods common in CCSN modeling. These moments, the energy density $E$ and flux $F_{j}$, must be related to the next highest moment, the pressure tensor $P_{ij}$. In the neutrino bulb model, in the absence of oscillations, we find that the $rr$-component of the pressure tensor can be related to the energy density via
\begin{equation}
    P_{rr} = \frac{(1-\cos^{3}\theta_\mathrm{max})}{3\,(1-\cos \theta_\mathrm{max})} E~.
\end{equation}
Using Equation~\ref{eq:costhetamax}, this becomes
\begin{equation}
    P_{rr} = \left(2 - \left(\frac{R_{\nu}}{r}\right)^{2} + \sqrt{1 - \left(\frac{R_{\nu}}{r}\right)^{2}}\right) \frac{E}{3}.
    \label{eq:2mom-closure}
\end{equation}
In what follows we shall refer to this equation as the ``two-moment`` closure. 
\subsubsection{Inner Boundary Condition}

To construct the initial values for the moments at the neutrinosphere we adopt a steady state distribution $f$ which is pure diagonal in the flavor basis. The diagonal entries of the distribution matrix are of the form $f^{(aa)}(R_{\nu},q,\theta) \propto Q_{a}(q)\,\Theta_{a}(\theta)$ with $Q_{a}$ a function describing the energy spectrum for flavor $a$ and $\Theta_{a}$ a function describing the angular spectrum for flavor $a$. For simplicity, the energy spectra are taken to be Fermi Dirac distributions
\begin{equation}
    Q_{a}(q) = \frac{1}{1+\exp{\left(q/(k_B\,T_{a}) - \eta_{a}\right)}} \label{eq:fermi-dirac}
\end{equation}
where $T_a$ is the temperature and $\eta_{a}$ is the chemical potential divided by $k_B T_a$. For both the moment-based and multi-angle calculations we adopt a uniform energy grid from $1\,\mathrm{MeV}$ to $60\,\mathrm{MeV}$ with 591 energy bins corresponding to an energy resolution of $100\;{\rm keV}$. Repeating all calculations using a resolution of $50\;{\rm keV}$ indicates the numerical error is less than 0.1\%. We adopt the parametric form for the initial angular distributions introduced by \citet{mirizzi_instability_2012}
\begin{equation}
\label{eq:angular_distribution}
\Theta_a(\theta) = \cos^{\beta_a}\theta.
\end{equation}
The case of half isotropic emission corresponds to $\beta_a = 0$. Using these distributions we write
\begin{equation}
f^{(aa)}(R_{\nu},q,\theta) = A_{a}\,(\beta_{a} + 2)\, Q_{a}(q)\,\Theta_{a}(\theta)
\end{equation}
and imposing the requirement that the luminosity matrix $L$ be given by equation  (\ref{eq:luminositymatrix}) at the neutrinosphere $R_{\nu}$, we find the constant $A_{a}$ to be
\begin{equation}
A_{a} = \frac{\pi}{G_{2}(\eta)}\left(\frac{c}{R_{\nu}}\right)^2\,\left(\frac{\hbar}{k_B T_{a}}\right)^3\, \frac{L^{(aa)}}{\langle q\rangle_{a}} \,\,.
\end{equation}
The function $G_{2}(\eta)$ is the complete Fermi-Dirac integral defined as 
\begin{equation*}
    G_{j}(x) = \int_{0}^{\infty} \frac{t^{j}}{1+\exp{\left(t - x\right)}} dt
\end{equation*}
and the quantity $\langle q \rangle_{a}$ is the mean energy given by
\begin{equation}
\langle q\rangle_{a} = \frac{ \int F^{(aa)}(R_{\nu},q) \,dq}{ \int F^{(aa)}(R_{\nu},q) /q\, dq}~.
\end{equation}
With the distribution matrix now defined, the energy density and flux moments at radius r (assuming no oscillations) are computed to be
\begin{widetext}
\begin{align}
E(r,q) &= \frac{R_{\nu}^2}{2\,r^2}\,\left(\frac{q}{2 \pi \hbar c}\right)^{3}
\begin{pmatrix} A_e\,Q_e(q)\,_{2}F_{1}\left(1,\frac{1}{2},2+\frac{\beta_e}{2},\frac{R_{\nu}^2}{r^2}\right) & 0 \\ 0 & A_{x}\,Q_{x}(q)\,_{2}F_{1}\left(1,\frac{1}{2},2+\frac{\beta_x}{2},\frac{R_{\nu}^2}{r^2}\right)
    \end{pmatrix} \\
F_r(r,q) & = \frac{R_{\nu}^2}{2\,r^2}\,\left(\frac{q}{2 \pi \hbar c}\right)^{3}
\begin{pmatrix} A_e\,Q_e(q) & 0 \\ 0 & A_{x}\,Q_{x}(q)
    \end{pmatrix} \\
P_{rr}(r,q) & = \frac{R_{\nu}^2}{2\,r^2}\,\left(\frac{q}{2 \pi \hbar c}\right)^{3}
\begin{pmatrix} A_e\,Q_e(q)\,_{2}F_{1}\left(1,-\frac{1}{2},2+\frac{\beta_e}{2},\frac{R_{\nu}^2}{r^2}\right) & 0 \\ 0 & A_{x}\,Q_{x}(q)\,_{2}F_{1}\left(1,-\frac{1}{2},2+\frac{\beta_x}{2},\frac{R_{\nu}^2}{r^2}\right)
    \end{pmatrix} 
\end{align}
\end{widetext}
where $_2F_1(a,b,c,z)$ is the ordinary hypergeometric function.
Similar expressions give the initial conditions for the antineutrino moments.

\subsection{Numerical Method}
To solve the quantum moment evolution equations we have developed a new code that computes the steady-state solution based on an inner boundary condition, given a specified  matter density and electron fraction throughout the computational domain. We omit the collision terms in the moment evolution equations in order to focus our attention upon the oscillation physics. The code solves the equations using an explicit midpoint (second order Runge-Kutta) integrator. The radial step is adaptive based on the three frequencies associated with the various flavor mixing terms in the Hamiltonian:
\begin{equation}
\begin{aligned}
    \omega_\mathrm{V} & = \frac{\Delta m_{12}^{2}\,c^4}{2\, \hbar\, q }  \\
    \omega_\mathrm{M} & = \sqrt{2}G_{F}\, n_{e}/\hbar \\
    \omega_\mathrm{SI} & = \sqrt{ ||H_{SI}||^2_{F} -\frac{1}{2}\,\left[ Tr(H_{SI}) \right]^2 }\,/ \hbar
    \label{eq:timestep}
\end{aligned}
\end{equation}
and $||H_{SI}||^2_{F}$ is the Frobenius norm of the self-interaction Hamiltonian, and $Tr(H_{SI})$ its trace. In the absence of the collision terms, the quantities $||F_{r}||^2_{F} - \frac{1}{2}\,\left[Tr(F_{r}) \right]^2$ and $||\bar{F}_{r}||^2_{F} - \frac{1}{2}\, \left[ Tr(\bar{F}_{r}) \right]^2$ are conserved, i.e.~ independent of the radius $r$, for every energy. Our code enforces the conservation of these quantities to an error tolerance set to $0.1\%$ per step. If the fractional change in the size of these quantities exceeds this bound, the time increment is halved and the step repeated. In what follows we present the results from several test problems of the code.

\section{Testing Moment-Based Methods\label{sec:tests}}

If moment-based approaches to flavor oscillations are to be a feasible alternative to calculations based on discrete ordinates or other more exact approaches, the results must agree well with analytic predictions if they exist and/or the results from less approximate numerical approaches. 

\subsection{Constant Electron Density\label{sec:constantdensity}}

Our first test case is flavor mixing of neutrinos emitted from a hard sphere of radius $R_{\nu}$ in a background of constant electron density. The flavor evolution for a single neutrino is well known for this scenario. The probability $P_{T}(r,\theta)$ that a neutrino initially in a particular flavor transitions to a neutrino of the opposite flavor after traveling a distance $\lambda$ is given by a sinusoid with fixed amplitude and wavelength. Specifically,
\begin{equation}
    P_T(r,\theta) = \sin^2\left(2\theta_\mathrm{MSW}\right) \sin^2\left(\frac{\omega_\mathrm{MSW}\, \lambda}{\hbar\,c}\right)~.
    \label{eq:P_transition}
\end{equation}
In this equation the effective matter mixing angle $\theta_\mathrm{MSW}$ and effective frequency $\omega_\mathrm{MSW}$, are
\begin{equation}
\begin{aligned}
\sin^{2} (2\,\theta_\mathrm{MSW}) & = \frac{\sin^2(2\,\theta_{12})}{\sin^2(2\,\theta_{12}) + C^2}\\
\omega_\mathrm{MSW} & = \frac{\omega_\mathrm{V}}{2} \sqrt{\sin^2(2\,\theta_{12}) + C^2} ~,
\end{aligned}
\label{eq:P_transition_MSW}
\end{equation}
where
\begin{equation}
C = \cos(2\theta_{12}) \mp \frac{\omega_\mathrm{M}}{\omega_\mathrm{V}}~.
\end{equation}
The negative sign is used for neutrinos and the positive sign for antineutrinos.

Using equation (\ref{eq:P_transition}), one can compute the angular moments of an ensemble of neutrinos by integrating the single neutrino solution over the solid angle with the appropriate $\cos \theta$ weight and taking into account the different path lengths $\lambda$ from the emission point to the radial point of interest. 
However one does not need to do this calculation to predict what one should observe in the solutions. 
Due to the differing path lengths, the neutrinos will not all have the same phase of the flavor oscillations (i.e., they will be not all be coherent). The incomplete coherence will inevitably smear the oscillations of the moments compared to the flavor oscillations of a single neutrino. 
The amount of coherence between the neutrinos on different trajectories is a function of the distance from the source. When the spread in path lengths to a given point is much smaller than the wavelength of the oscillations, all the neutrinos are very close to being coherent, but as one moves away from the source the spread in path lengths grows and the amount of coherence drops. For a neutrinosphere of radius $R_{\nu}$, the path length difference between neutrinos emitted radially and those emitted at a tangent to the neutrinosphere asymptotes to $R_{\nu}$. Thus the amount of coherence should also asymptote to a level determined by the ratio of the oscillation wavelength to $R_{\nu}$. 

In order to test our code's ability to reproduce these predictions we solve the two moment evolution equations and separately perform a multi-angle calculation using SQA. The spatial grid extends to $40\,\mathrm{km}$, we use artificial vacuum mixing parameters of $\Delta m^{2}_{12} = 6.9 \times 10^{-4}\, \mathrm{eV}^{2}$ and $\theta_{12} = 0.28818$ and consider monoenergetic neutrinos with an energy of $q= 1\,\mathrm{MeV}$. The elements of the energy density moments are set arbitrarily according to the hierarchy $E^{(ee)}(R_{\nu}) = 10\, \bar{E}^{(ee)}(R_\nu) = 100\, E^{(xx)}(R_\nu) = 100\, \bar{E}^{(xx)}(R_\nu)$. The matter density is set to $\rho=8\times10^3\,\mathrm{g\,cm}^{-3}$ and the electron fraction is $Y_e=0.5$ in order to put the $1\,\mathrm{MeV}$ neutrinos close to the MSW resonance.  

\begin{figure}[t]
\centering
\includegraphics[width = 0.5\textwidth]{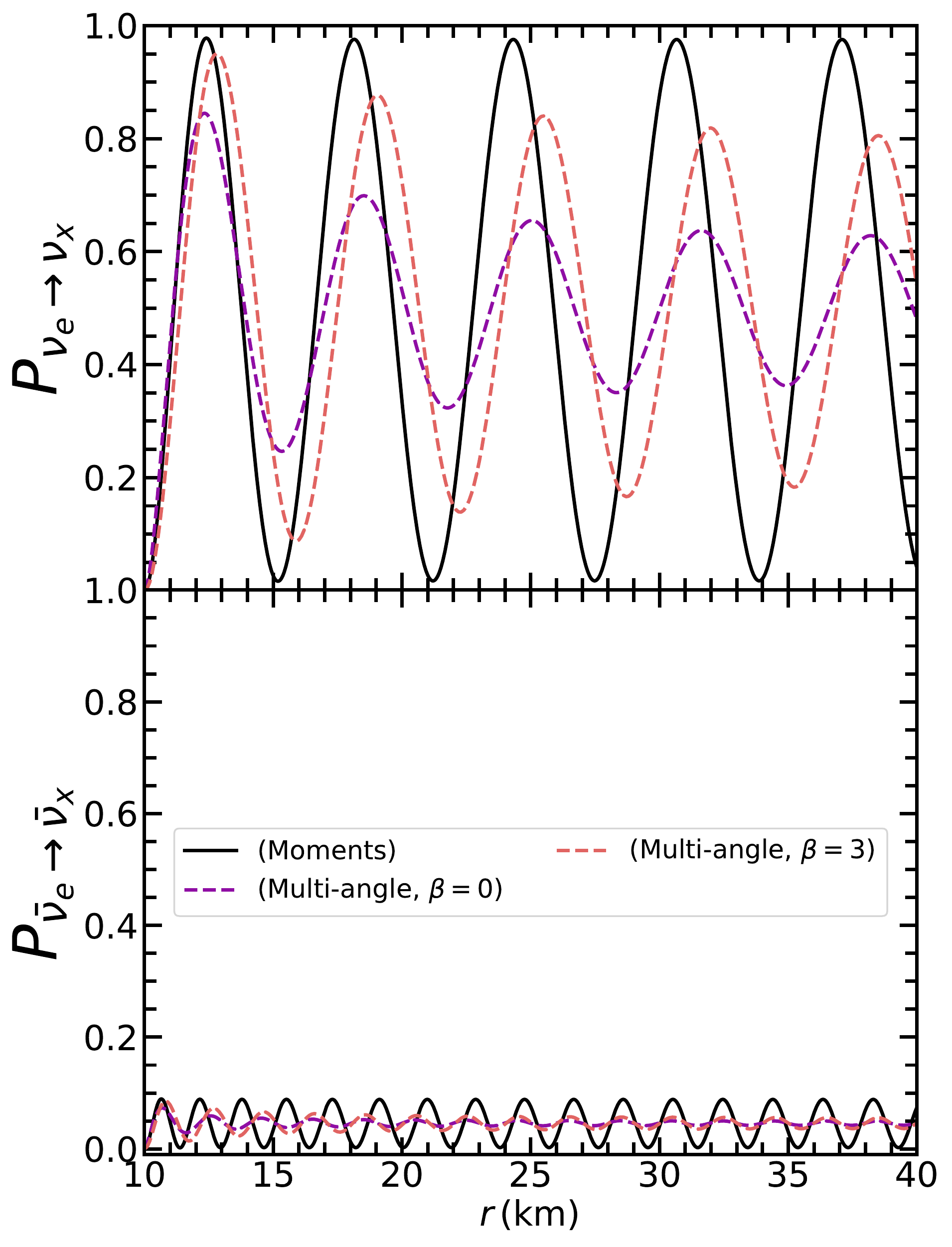}
\caption{Flavor oscillation test including vacuum and matter contributions to the Hamiltonian. Here we use $\Delta m^{2}_{12} = 6.9 \times 10^{-4}\, \mathrm{eV}^{2}$ and $\theta_{12} = 0.28818$ in a background density profile of $\rho = 8000 \,\mathrm{g/cm}^{3}$ and $Y_{e} = 0.5$. The top panel shows the transition probability of electron neutrinos to $x$-flavor neutrinos while the bottom is the same for the antineutrinos. Solid lines are the results from the moment code and dashed lines are the results from the SQA using 90,001 angular bins.}
\label{fig:constantdensity_test}
\end{figure}

Figure~\ref{fig:constantdensity_test} shows the oscillation probabilities as a function of the radius $r$ for two angular distribution parameters of ${\beta_a} = 0$ (dashed red) and ${\beta_a}=3$ (dashed purple). 
The electron neutrino and antineutrino transition probabilities are defined from the flux moment to be
\begin{eqnarray}
P_{\nu_e \rightarrow \nu_x}(r) & = & \frac{r^2\,F_{r}^{(xx)}(r) - R_{\nu}^2\,F_{r}^{(xx)}(R_{\nu})}{R_{\nu}^2\,[F_{r}^{(ee)}(R_{\nu}) - F_{r}^{(xx)}(R_{\nu})]} \label{eq:Ptrans}\\
{\bar P}_{\bar{\nu}_e \rightarrow \bar{\nu}_x}(r) & = & \frac{r^2\,\bar{F}_{r}^{(ee)}(r) - R_{\nu}^2\,\bar{F}_{r}^{(xx)}(R_{\nu})}{R_{\nu}^2\,[ \bar{F}_{r}^{(ee)}(R_{\nu}) - \bar{F}_{r}^{(xx)}(R_{\nu})] } \label{eq:Ptransbar}.
\end{eqnarray}
The top panel shows the flavor evolution as a function of radius for the neutrinos and the bottom panel for the antineutrinos. Although the neutrinos are on resonance and undergo nearly complete flavor oscillations, antineutrinos are off-resonance and only undergo minor flavor transformation. A comparison of the SQA results for the half-isotropic case and a separate, analytic integration of the survival probability, i.e., 
\begin{equation}
P_{\nu_e \rightarrow \nu_x}(r) = \frac{2\,r^2}{R_{\nu}^2}\,\int^1_{\cos\theta_{max}} P_T(r,\theta)\,\cos\theta\,d(\cos\theta)\,\,,
\end{equation}
are visually indistinguishable. 

The SQA results (dashed) display the decoherence effect of the neutrinos that have traveled along different trajectories leading to the reduction in the amplitude of the oscillations with increasing radius. Close to the neutrinosphere, the amplitudes of the flavor oscillations predicted from the moment code (solid) are similar in magnitude to the ${\beta_a}=0$ SQA results (dashed purple). However, as the neutrinos move away, the moment code maintains larger flavor oscillations than exhibited in the multi-angle results.

Although the moment code overestimates the coherence of the neutrinos traveling along different trajectories, it does capture some phase effects. As the distribution becomes more forward-peaked, the average phase advances more slowly due to smaller average path length to a given radial point. This can be seen by comparing the SQA results for ${\beta_a}=3$ (forward-peaked) and ${\beta_a}=0$ (semi-isotropic). The moment results, which have initial conditions corresponding to ${\beta_a}=0$, show a more slowly evolving phase than the ${\beta_a}=3$ results, just as the SQA ${\beta_a}=0$ results do. However, the agreement does not last for long and a significant phase difference between the the moment and ${\beta_a}=0$ oscillations exists at larger radius. Of course, in the limit of a perfectly forward-peaked distribution (not shown), the moment equations become increasingly accurate and agree increasingly well with multi-angle data.


\begin{figure}[t]
\centering
\includegraphics[width = 0.5\textwidth]{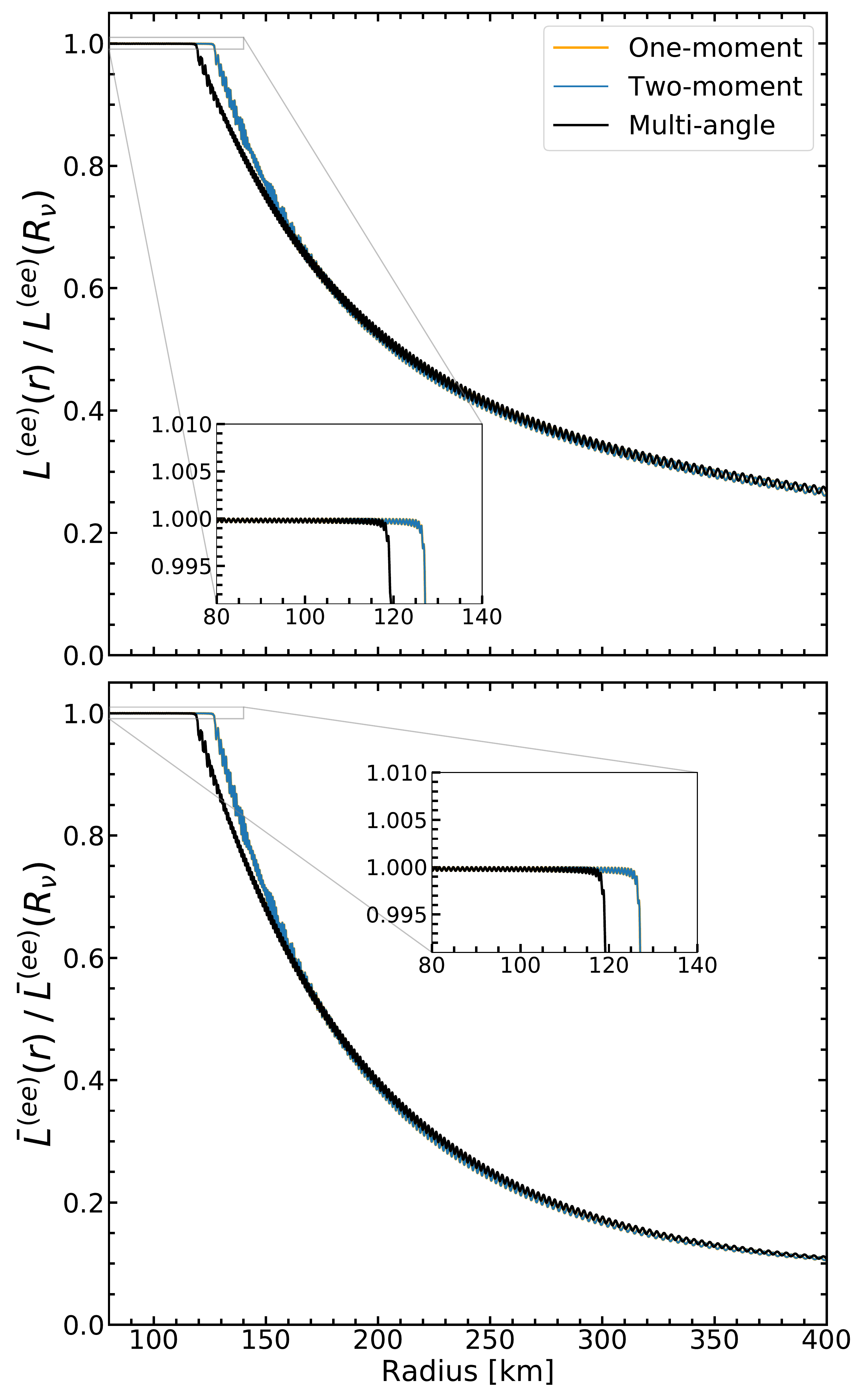}
\caption{The ratio of the `ee' element of luminosity matrix relative to its initial value, versus radius for the test in Section~\ref{sec:collective1} and Table~\ref{tab:collective_setup_tableI}. The top panel is for neutrinos and the bottom panel is for antineutrinos. The one-moment results are difficult to discern due to being largely obscured by the two-moment results.}
\label{fig:luminosity_table1}
\end{figure}

\subsection{The Steady State Self Interaction Problem In Spherical Symmetry}

Our next three tests are for the case of collective neutrino oscillations in an inhomogeneous environment. Collective flavor mixing is of interest for the study of core-collapse supernovae, as it is thought that such oscillations may occur at late times above the protoneutron star once the explosion has occurred. This may have a dramatic impact on the observed neutrino signal and nucleosynthesis in the CCSN environment \cite{duan_collective_2010}. Collective oscillations are well-studied in the steady state, free streaming approximation \cite{2020arXiv201101948T}. One of the distinct features of collective flavor mixing are the spectral swaps and splits that arise in the neutrino and antineutrino distributions at large radii (e.g., \cite{duan_neutrino_2007,dasgupta_multiple_2009}). This feature causes the neutrino spectra to deviate from the thermal-like spectra emitted from the neutrinosphere. It is also well established from such studies that the outcome of these calculations is heavily dependent on fully resolving the angular distribution of neutrinos \cite{2012PhRvD..86l5020S}. This makes it a challenging test of a moment-based approach.

We present three simulations of this system using both the one-moment and two-moment closures in the following subsections. The first calculation in Section~\ref{sec:collective1} is a reasonably realistic representation of the conditions outside a protoneutron star after the onset of explosion. The second calculation in Section~\ref{sec:collective2} is a modified version of this same setup, where we artificially adjust the parameters in order to probe the behavior of the moment method when flavor transformations occur in regions where the neutrino distribution is not highly forward-peaked. Finally, the third calculation in Section~\ref{sec:collective3} is the interesting and demanding case of a double spectral split which tests the ability of moments to track the correct degree of coherence in the neutrinos. 


\begin{table}[t]
    \centering
    \begin{tabular}{l| c c c c c c c}
        a & $L_a$ [ergs/s] & $\langle q \rangle_a$ [MeV] & $T_a$ [MeV] & $\eta_a$ & ${\beta_a}$ & $F_r/E$ & $P_{rr}/E$ \\
        \hline 
    $\nu_{e}$ & $4.1\times10^{52}$ & 9.4 & 2.1 & 3.9 & 0.0 & 0.9975 & 0.995 \\
    $\bar{\nu}_{e}$ & $4.3\times 10^{52}$ & 13.0 & 3.5 & 2.3 & 0.0 & 0.9975 & 0.995 \\
    $\nu_{x}$ & $3.95 \times 10^{51}$ & 15.8 & 4.4 & 2.1 & 0.0 & 0.9975 & 0.995 \\
    $\bar{\nu}_{x}$ & $3.95\times10^{51}$ & 15.8 & 4.4 & 2.1 & 0.0 & 0.9975 & 0.995
    \end{tabular}
    \caption{Parameters for the realistic collective oscillation simulation in Section~\ref{sec:collective1}. Listed are the neutrino luminosity $L_a$, average energy $\langle q \rangle_a$, temperature $T_a$, and chemical potential divided by the temperature $\eta_a$ used in Equation~\ref{eq:fermi-dirac}.  These are modified from the case of $p=10$ and $q=3.5$ (different from the $q$ in this work that represents neutrino energy) from Table~6 in \citet{keil_monte_2003}, with the electron flavor luminosities increased by a factor of 10. All angular distributions are described by $\beta_a=0$ in Equation~\ref{eq:angular_distribution}. The initial flux factors $F_r/E$ and initial Eddington factors $P_{rr}/E$ are also provided.}
    \label{tab:collective_setup_tableI}
\end{table}

\begin{figure*}[t]
\centering
\includegraphics[width =0.9 \textwidth]{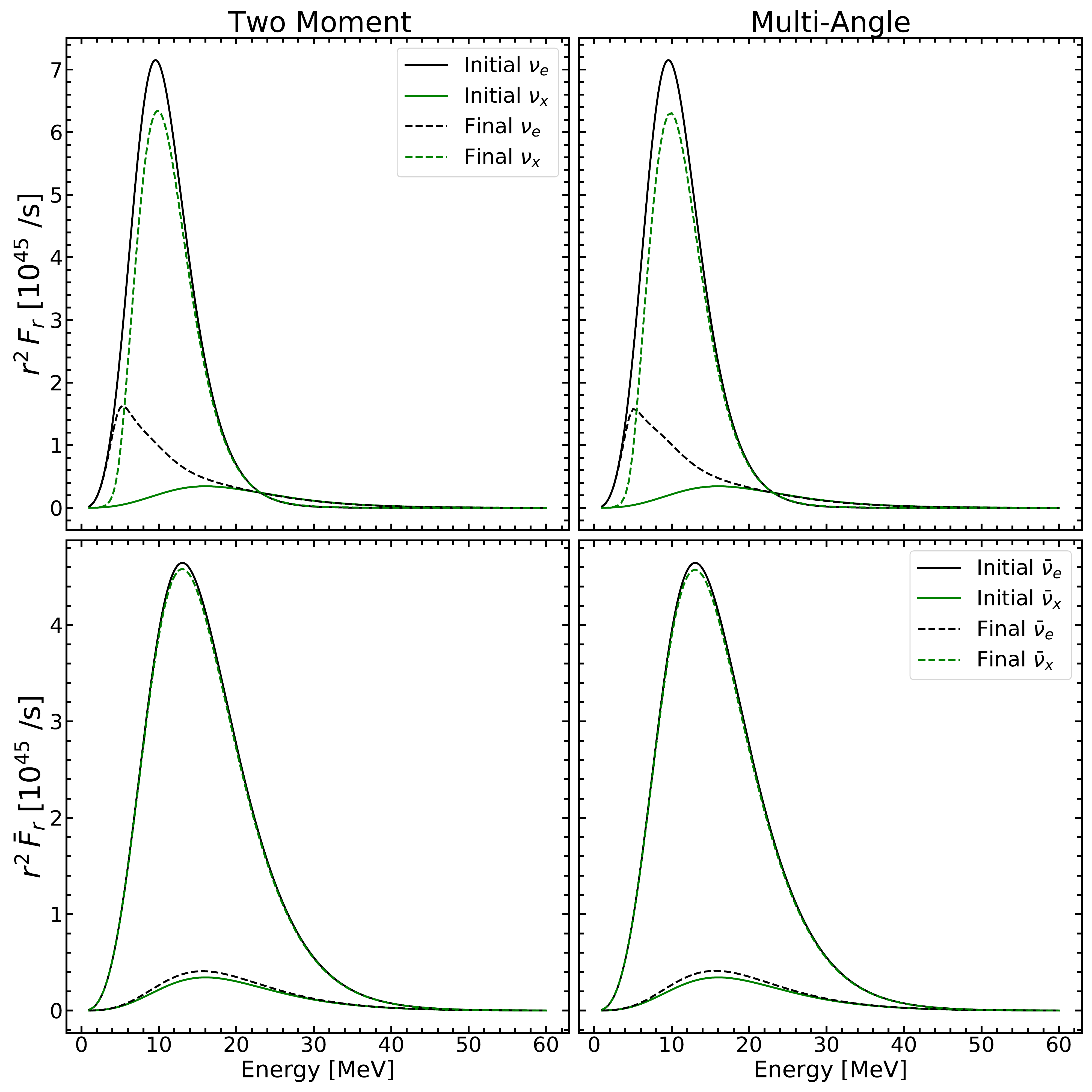}
\caption{Initial and final spectra versus energy for the test in Section~\ref{sec:collective1} and Table~\ref{tab:collective_setup_tableI}.  Neutrinos are in the top two panels and antineutrinos are in the bottom two.  The left column shows the two-moment simulation and the right column shows the multi-angle simulation. The black curves are the electron flavor and the green curves are the $x$ flavor.  The solid curves show the initial spectra  and the dashed curves show the final spectra.  The neutrino spectrum in all cases shows the spectral splits common in collective mixing. The antineutrinos show an almost complete spectral swap. }
\label{fig:spectrum_tableI}
\end{figure*}

\subsubsection{Instability far from the neutrinosphere}
\label{sec:collective1}

In Table~\ref{tab:collective_setup_tableI} we present the values for the neutrino luminosity $L_{a}$, average energy $\langle q \rangle_{a}$, temperature $T_{a}$, and chemical potential $\eta_{a}$ to be used in the first collective test calculation. These values are a modified version of the case $p=10$ and $q=3.5$ in Table~6 of \citet{keil_monte_2003}, with the electron neutrino and antineutrino luminosities increased by a factor of 10. This makes the electron neutrino and electron antineutrino luminosities typical of what one expects during the accretion phase of an iron-core collapse supernova - see, for example, figure 2 in \cite{2010A&A...517A..80F}. The inner computational boundary was chosen to be $100\,\mathrm{km}$ and the system was allowed to evolve until the neutrinos reached the radius of $400\,\mathrm{km}$. We set the mixing parameters to $\Delta m^{2}_{12} = -2.7 \times 10^{-3}\;{\rm eV}^2$ (similar to $\Delta m_{32}^{2}$ in the Inverted Mass ordering \cite{Zyla:2020zbs}) and $\theta_{12} = 0.01$ in order to emulate the effect of matter suppression. The neutrinosphere is set to $R_{\nu} = 10\,\mathrm{km}$ for all neutrino and antineutrino flavors. 

In Figure ~\ref{fig:luminosity_table1} we show the results of the calculation. The figure shows The ratio of the `ee' element of luminosity matrix relative to its initial value as a function of the radius. 
In the figure we see the moment calculations exhibit a flavor instability at $126\,\mathrm{km}$, similar to, but slightly larger than the radius of $119\,\mathrm{km}$ at which the flavor instability is seen in the multi-angle simulation. For both approaches, once the instability has begun we see that antineutrinos experience an almost complete flavor swap. 
The figure also indicates that there is little difference between the moment calculations that use the one-moment and two-moment closures. This similarity was not enforced nor expected and motivated the next test problem in Section~\ref{sec:collective2}. 

There is also remarkable agreement between the flavor-transformed spectra between the moment and multi-angle methods. Figure~\ref{fig:spectrum_tableI} shows the initial and final spectra from the one-moment, two moment, and multi-angle calculations. In both the moment-based and multi-angle calculations, the antineutrino spectra show an almost complete swap between the $\bar{\nu}_{e}$ and $\bar{\nu}_{x}$ flavors and the neutrinos show a split in the spectra at about $25\,\mathrm{MeV}$. This phenomenon has been seen many times in previous studies starting with Duan \emph{et al.} \cite{duan_collective_2006,PhysRevLett.103.051105}.

\subsubsection{An instability close to the neutrinosphere}
\label{sec:collective2}

\begin{table}[t]
    \centering
    \begin{tabular}{l| c c c c c c c}
        a & $L_a$ [ergs/s] & $\langle q \rangle_a$ [MeV] & $T_a$ [MeV] & $\eta_a$ & ${\beta_a}$ & $F_r/E$ & $P_{rr}/E$ \\
        \hline 
    $\nu_{e}$ & $2.050\times10^{49}$ & 9.4 & 2.1 & 3.9 & 0.0 & 0.5 & 0.33 \\
    $\bar{\nu}_{e}$ & $2.550\times 10^{49}$ & 13.0 & 3.5 & 2.3 & 0.0 & 0.5 & 0.33 \\
    $\nu_{x}$ & $1.698 \times 10^{49}$ & 15.8 & 4.4 & 2.1 & 0.0 & 0.5 & 0.33\\
    $\bar{\nu}_{x}$ & $1.698\times10^{49}$ & 15.8 & 4.4 & 2.1 & 0.0 & 0.5 & 0.33
    \end{tabular}
    \caption{Parameters for the collective oscillation simulation discussed in Section~\ref{sec:collective2} designed to expose independent evolution of multiple moments. Listed are the luminosity $L_a$, average energy $\langle q \rangle_a$, temperature $T_a$, chemical potential divided by the temperature $\eta_a$, the angular distribution parameter $\beta_a$, the initial flux factor $F_r/E$ and initial Eddington factor and $P_{rr}/E$. Due to the rapid increase of the run time of the multi-angle calculation as the inner boundary is moved towards the neutrinosphere, the multi-angle calculation was started at 1 km above the neutrinosphere.}
    \label{tab:setup_tableII}
\end{table}

As we pointed out in Section~\ref{sec:collective1}, the high degree of concordance between the results using the one and two-moment closures was not enforced nor expected, but in hindsight is perhaps not surprising. At radii well beyond the neutrinosphere, there is very little difference between the flux and pressure moments, since the flux and Eddington factors are close to unity. One cannot expect to resolve fine angular features in a pencil-beam distribution using only coarse moments. If this interpretation is correct, we might reasonably expect to observe more significant differences between the one and two-moment calculations when flavor instabilities occur closer to the neutrinosphere (i.e., where the flux and Eddington factors are not close to unity). For this reason we present results from a second self-interaction test case where we engender flavor transformation closer to the neutrinosphere by artificially adjusting the neutrino luminosities. The new set of parameters we use for this calculation are shown in Table~\ref{tab:setup_tableII}. In addition, for this test the inner boundary of the computational domain is at the neutrinosphere ($R_\nu=10\,\mathrm{km}$). 

\begin{figure}[t]
\centering
\includegraphics[width = 0.5\textwidth]{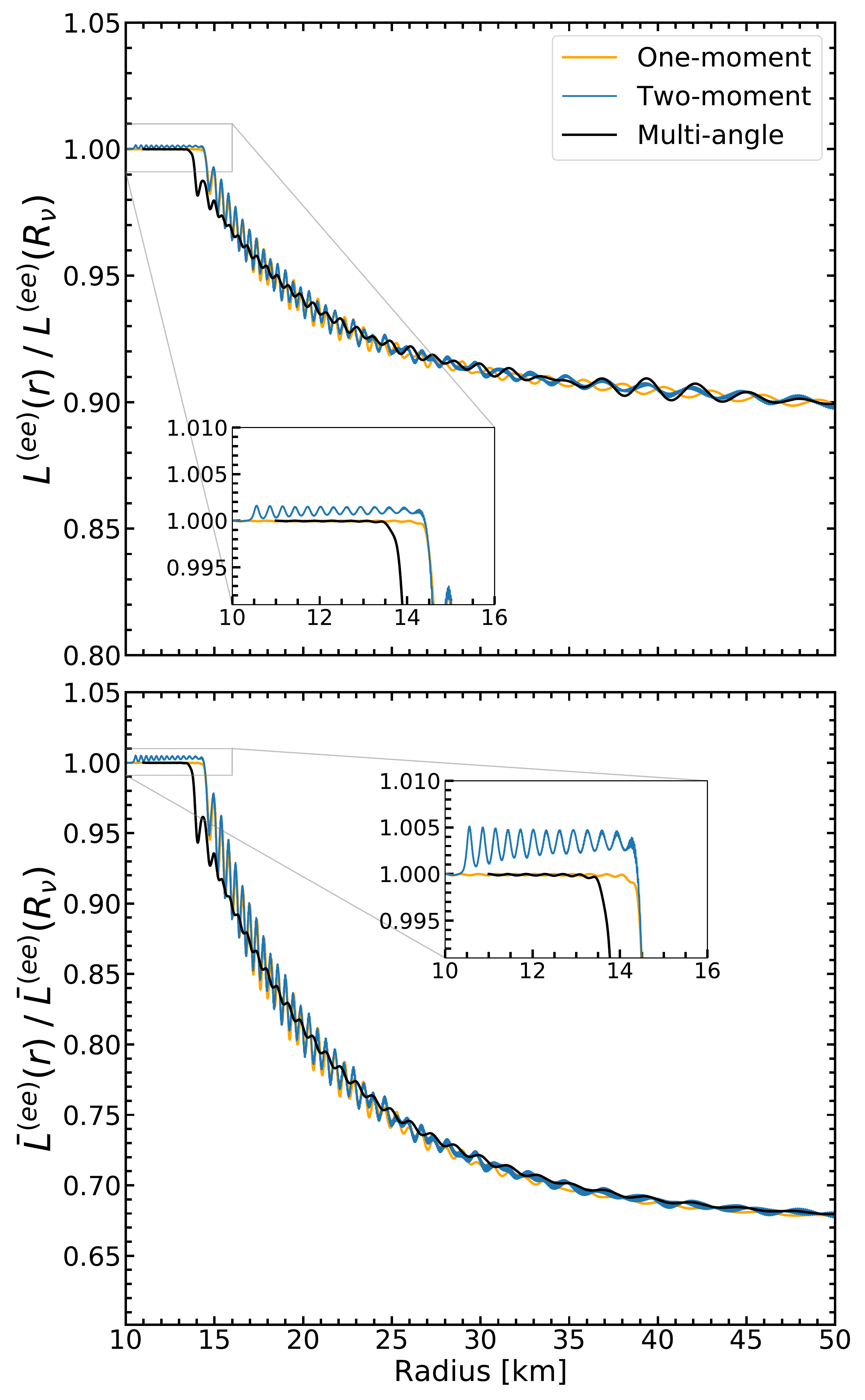}
\caption{The ratio of the `ee' element of luminosity matrix relative to its initial value, for the modified collective oscillation test in Section~\ref{sec:collective2} and Table~\ref{tab:setup_tableII}. The top panel is for neutrinos and bottom panel is for antineutrinos. The orange curve shows the one-moment simulation, blue curve shows the two-moment simulation, and the black curve shows the multi-angle simulation.}
\label{fig:luminosity_tableII}
\end{figure}

The evolution of the `ee' element of luminosity matrix relative to its initial value from this second self-interaction calculation are shown in Figure~\ref{fig:luminosity_tableII}. The system behaves similarly to the previous test problem studied in Section~\ref{sec:collective1} in that the moment calculations (blue and orange) closely track each other and exhibit an onset of flavor transformation in approximately the same location as in the multi-angle calculations. However, due to the decreased neutrino luminosities, the instability starts at $r\approx14\,\mathrm{km}$, much closer to the neutrinosphere than the onset at $r\approx120\,\mathrm{km}$ in Section~\ref{sec:collective1}. The moment calculations agree with the multi-angle calculations about the onset of the instability even better than in the previous case. 

In the insets in Figure~\ref{fig:luminosity_tableII}, we show the same ratio of the ratio of the `ee' element of luminosity matrix relative to its initial value, during the first five kilometers of the calculation. We observe that the ratios from the multi-angle and one-moment calculations are small and slightly below unity over this region, but that the two-moment calculation are slightly above unity until significant flavor conversion occurs at $14\,\mathrm{km}$. Our investigation into the origin of the greater-than-unity luminosities - which are not physical - are presented in Appendix \ref{sec:NumericalErrors}. These investigations revealed that the origin is not numerical error and are likely due to the closure. 

The similarity between the results for the one- and two-moment closures shown in Figure~\ref{fig:luminosity_tableII} cannot be attributed to the fact that the flux and Eddington factors are close as in Section~\ref{sec:collective1}. Instead, we have found that this behavior arises because our choice of closure is so self-consistent that the evolution equation for the flux (Equation~\ref{eq:energy}) is essentially identical using both the one-moment and two-moment closures. Thinking first about the analytic relationships for the moments in a bulb model without flavor transformation, one can relate the pressure to the flux as $P_{rr}=D(r)\,F_r$, where $D(r)$ is a scalar function given by
\begin{equation}
\begin{aligned}
D(r) &= \frac{2\,(1+\cos\theta_{max}+\cos^2\theta_{max})}{3\,(1+\cos\theta_{max})}, \\
&= \frac{2}{3\,r} \left( \frac{2\,r^2-R^2_\nu + r\,\sqrt{r^2-R^2_\nu}}{r + \sqrt{r^2-R^2_\nu}} \right).
\end{aligned}
\label{eq:P_F_relationship}
\end{equation}
Inserting this relationship into Equations~\ref{eq:energy} and \ref{eq:flux}, the evolution equation for the steady-state flux in the two-moment approach becomes the same as the radial evolution of the steady state flux using the one-moment closure, indicating the self-consistency of our choice of closures. Thus, wherever the flux and pressure as calculated by the two-moment approach are related by Equation~\ref{eq:P_F_relationship}, the evolution of the flux in the one- and two-moment systems will evolve identically. 

\begin{figure}[t]
    \centering
\includegraphics[width=\linewidth]{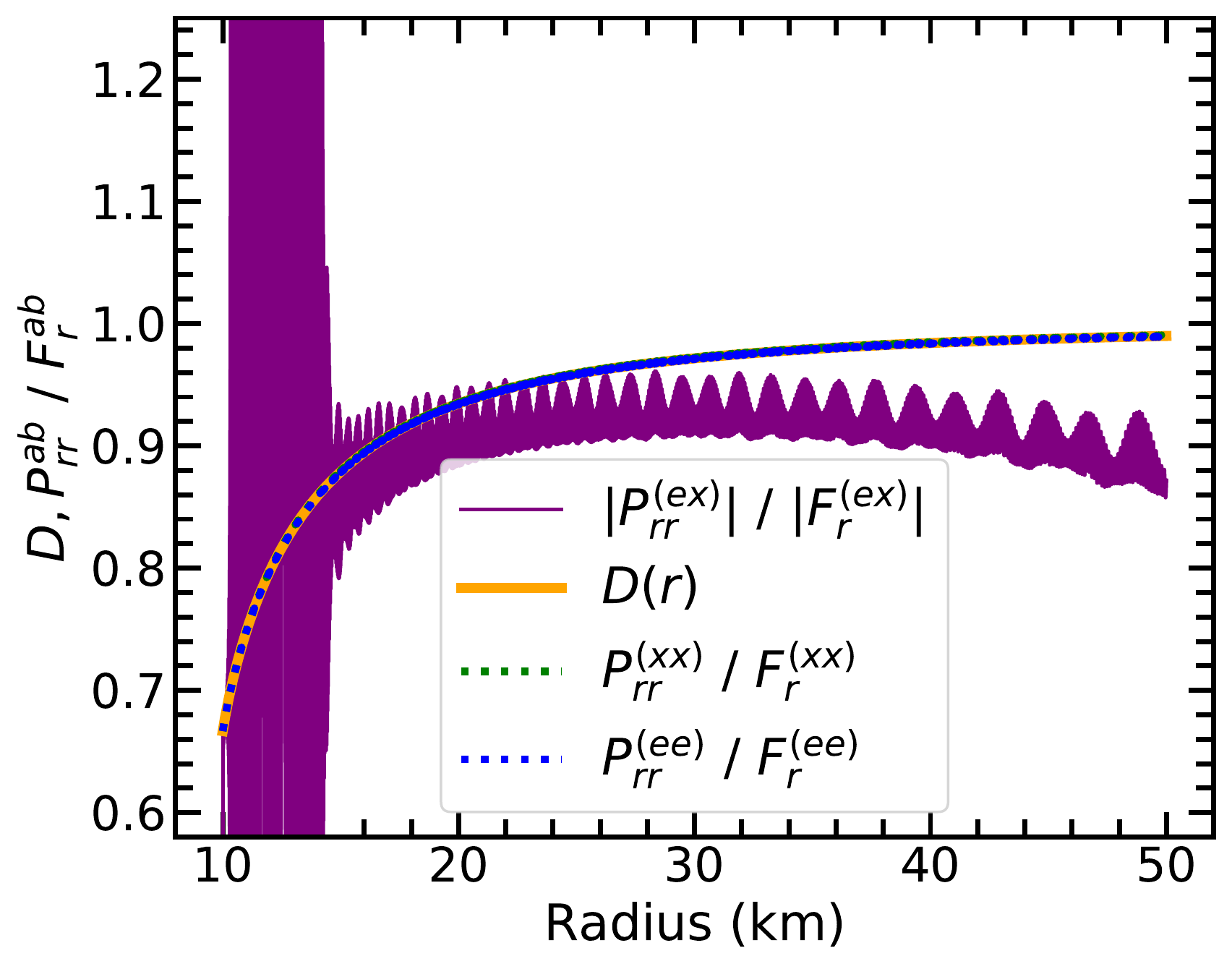}
\caption{The ratio of the elements from the pressure tensor to the corresponding elements from the flux as a function of the radius $r$ for the $15\;{\rm MeV}$ neutrinos in the two-moment calculation using the parameters given in Section~\ref{sec:collective2} and Table~\ref{tab:setup_tableII}. }
\label{fig:ratio}
\end{figure}

\begin{figure}[h]
    \centering
    \includegraphics[width=\linewidth]{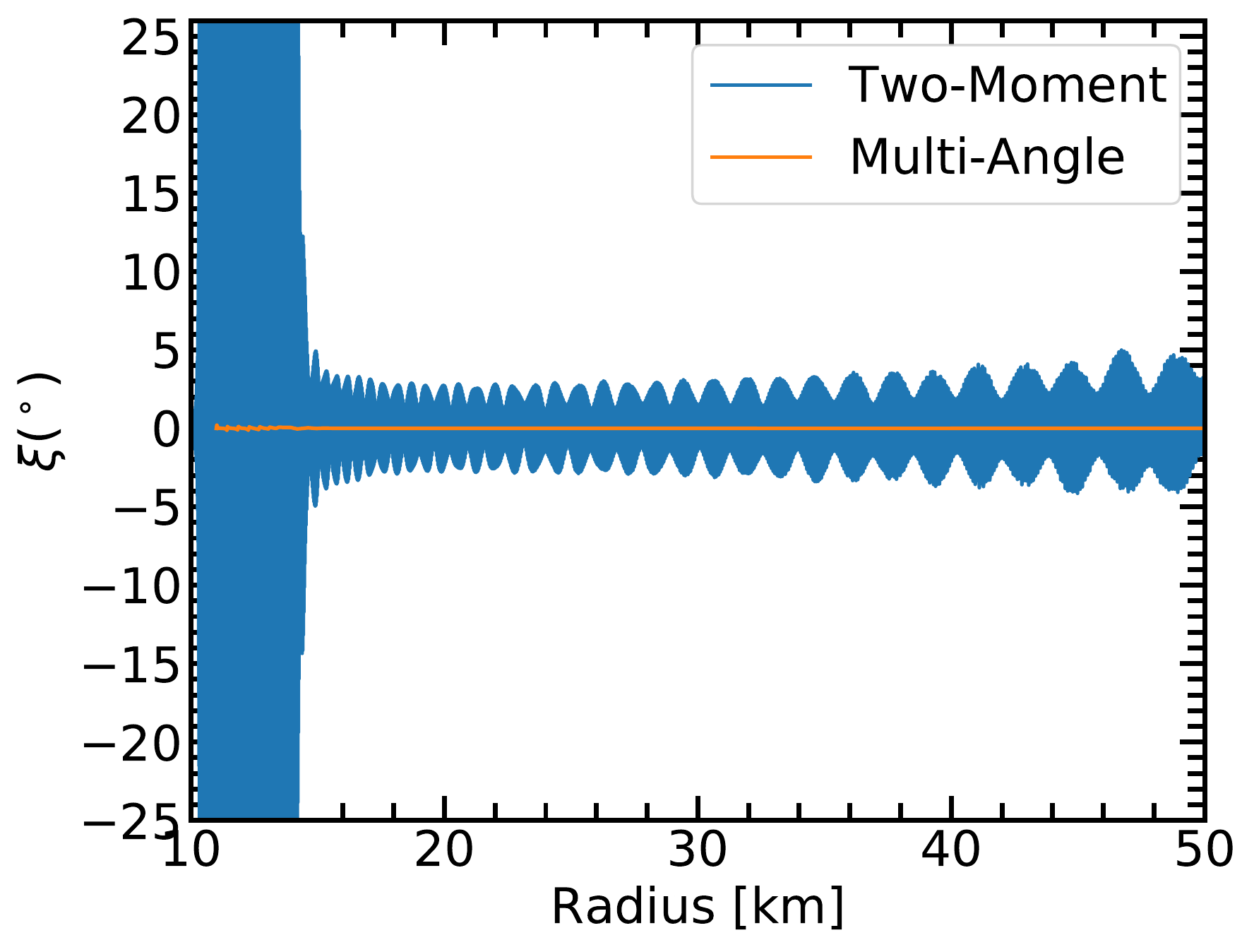}
    \caption{The difference between the phase angles of the off-diagonal elements of the pressure and flux moments for 15 MeV neutrinos from the two-moment simulation (cyan) and the multi-angle simulation (orange) in the modified collective oscillation setup of Section~\ref{sec:collective2} and Table~\ref{tab:setup_tableII}.}
    \label{fig:alignment}
\end{figure}

To examine whether this coherence of the flux and pressure moment actually occurs in our calculations, in Figure \ref{fig:ratio} we plot $D(r)$ (orange) together with the ratio of the corresponding elements from the pressure and flux moments for the  $15\;{\rm MeV}$ neutrinos. We find the ratios of $P_{rr}^{(ee)} / F_{r}^{(ee)}$ (blue dashed) and $P_{rr}^{(xx)}/F_{r}^{(xx)}$ (green dashed, not visible under the blue dashed curve) match the function $D(r)$ very well, indicating that the diagonal moments grow and shrink due to flavor transformation at the same rate. The ratio $|P_{rr}^{(ex)}|/|F_{r}^{(ex)}|$ (purple) does not follow $D(r)$ before the instability, as the flavor off-diagonal elements are very small. Once flavor transformation begins at $r\sim 14\;{\rm km}$, the flavor off-diagonal ratio more closely follows $D(r)$, though not as well as the diagonal elements. 

Figure~\ref{fig:alignment} shows the phase angle difference $\xi = \mathrm{arg}(P_{rr}^{(ex)}) - \mathrm{arg}(F_{r}^{(ex)})$ between the off-diagonal elements of the flux and pressure moments from the two-moment and multi-angle calculations. The figure shows how the phase difference for the multi-angle calculation is always small, indicating that the flux and pressure moments experience tightly coupled evolution. The phase difference in the two-moment calculation is initially highly variable, since the off-diagonal components are very small. However, once the flavor transformation begins around $r=14\;{\rm km}$, the two moments become very coherent and the phase difference does not exceed a few degrees. Taken together, Figures \ref{fig:ratio} and \ref{fig:alignment} show that $P_{rr}=D(r)\,F_r$ is a good approximation for this problem and the coherence between the two moments reduces the system of moment equations to that for a single moment. Note that this coherence between the pressure and flux in the two-moment scheme is not imposed, it emerges as the calculation proceeds.

\begin{figure*}[t]
\centering
\includegraphics[width =0.9 \linewidth]{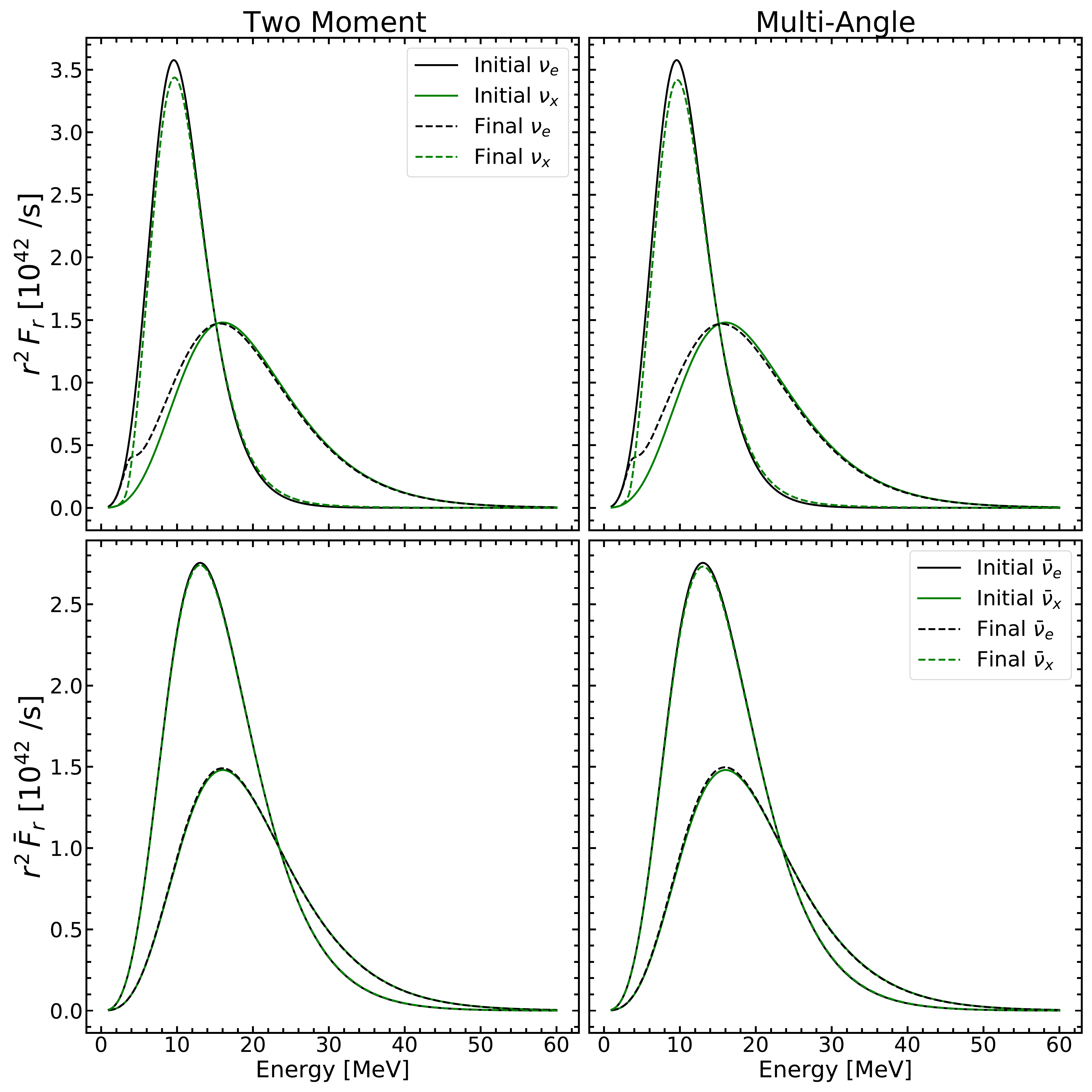}
\caption{Initial and final spectra versus energy for neutrinos from simulations using parameters in Table~\ref{tab:setup_tableII} and Section~\ref{sec:collective2}.  Neutrinos are in the top two panels and antineutrinos are in the bottom two.  The left column shows the two-moment simulation and the right column shows the multi-angle simulation. The black lines are the electron flavor and the green lines are the $x$ flavor. The solid lines show the initial spectra and the dashed lines show the final spectra.}
\label{fig:spectrum_tableII}
\end{figure*}

Finally, Figure~\ref{fig:spectrum_tableII} shows the spectra of the neutrinos and antineutrinos at $50\;{\rm km}$ from the two-moment and multi-angle calculations. The spectra from the two methods agrees well with only slight differences in the high-energy tails. The moment-based approach accurately calculates the spectral evolution as well as the gross, integrated behavior.


\begin{figure}[t]
\centering
\includegraphics[width = 0.5\textwidth]{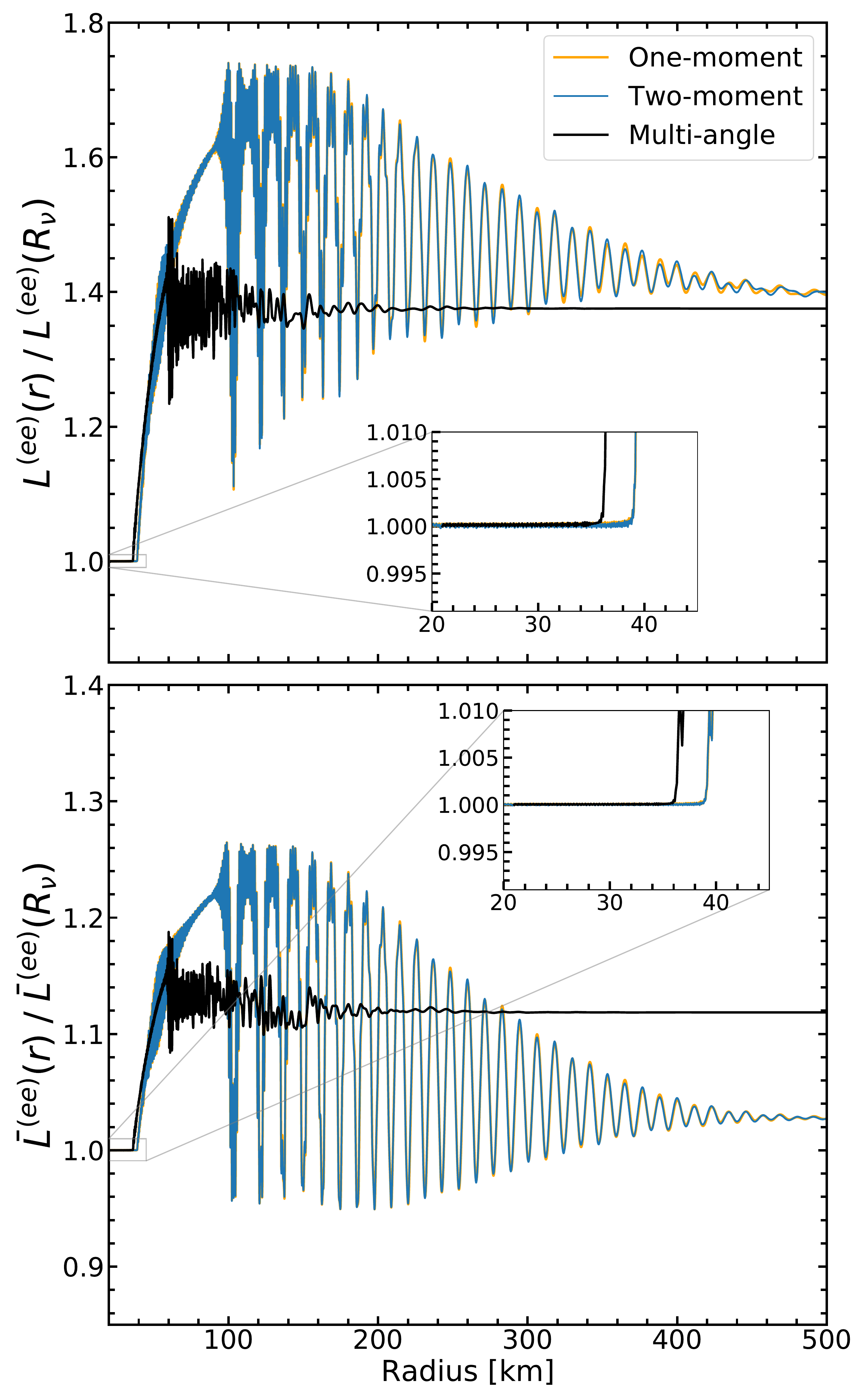}
\caption{The ratio of the `ee' element of luminosity matrix relative to its initial value, for the modified collective oscillation test in Section~\ref{sec:collective3} and Table~\ref{tab:setup_tableIII}. The top panel is for neutrinos and bottom panel is for antineutrinos. The orange curve shows the one-moment simulation, blue curve shows the two-moment simulation, and the black curve shows the multi-angle simulation.}
\label{fig:luminosity_tableIII}
\end{figure}

\begin{figure}[t]
\centering
\includegraphics[width =0.5 \textwidth]{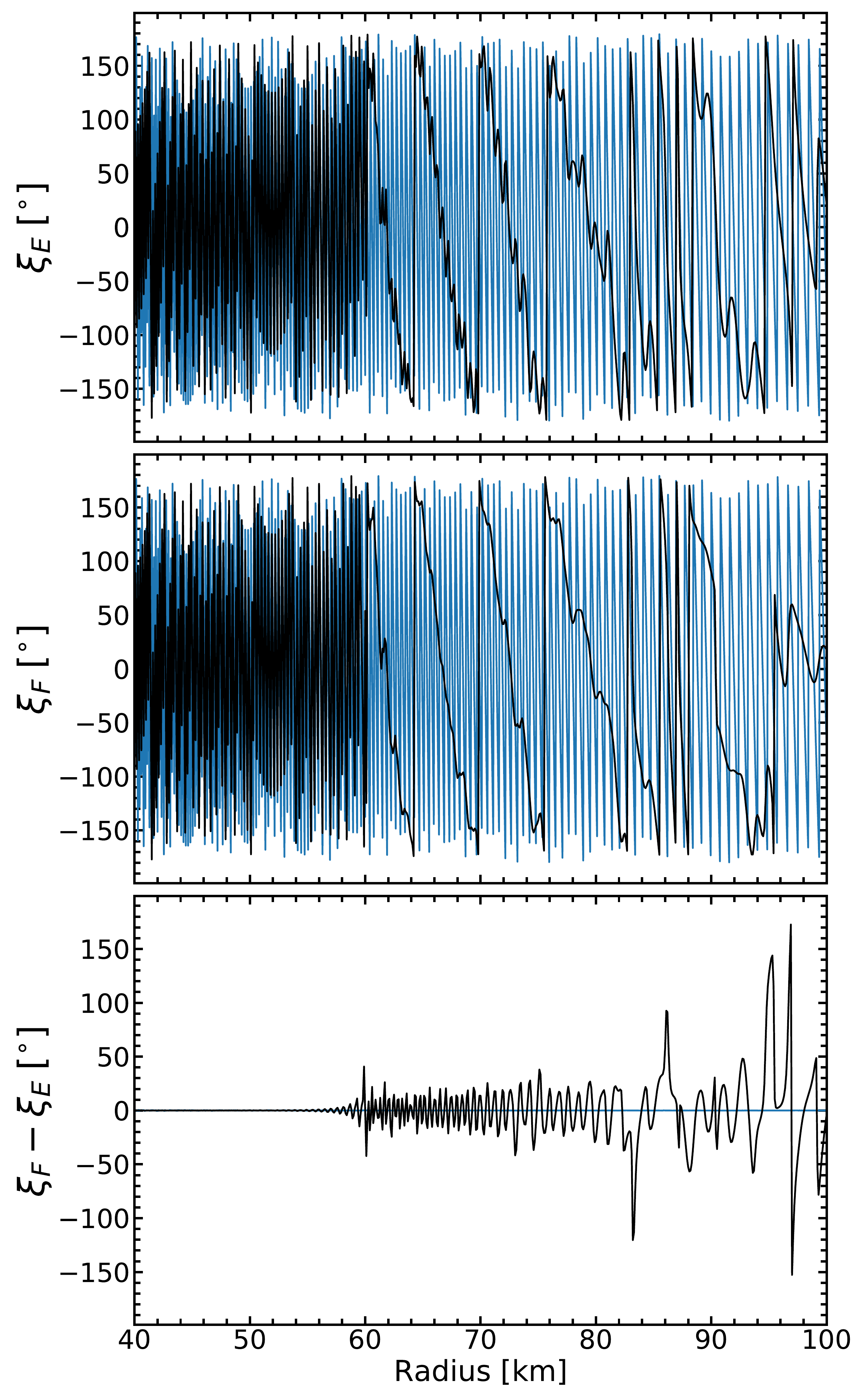}
\caption{The phase of the off-diagonal element of the $H_E$ component of the Hamiltonian (top panel), the $H_F$ contribution (middle panel) and the difference between them (bottom panel) as a function of the radial coordinate. Blue lines are the results from the moment-based approach using the two-moment closure, the black are from the multi-angle calculation.}
\label{fig:T3phases}
\end{figure}

\begin{figure*}[t]
\centering
\includegraphics[width =0.9 \linewidth]{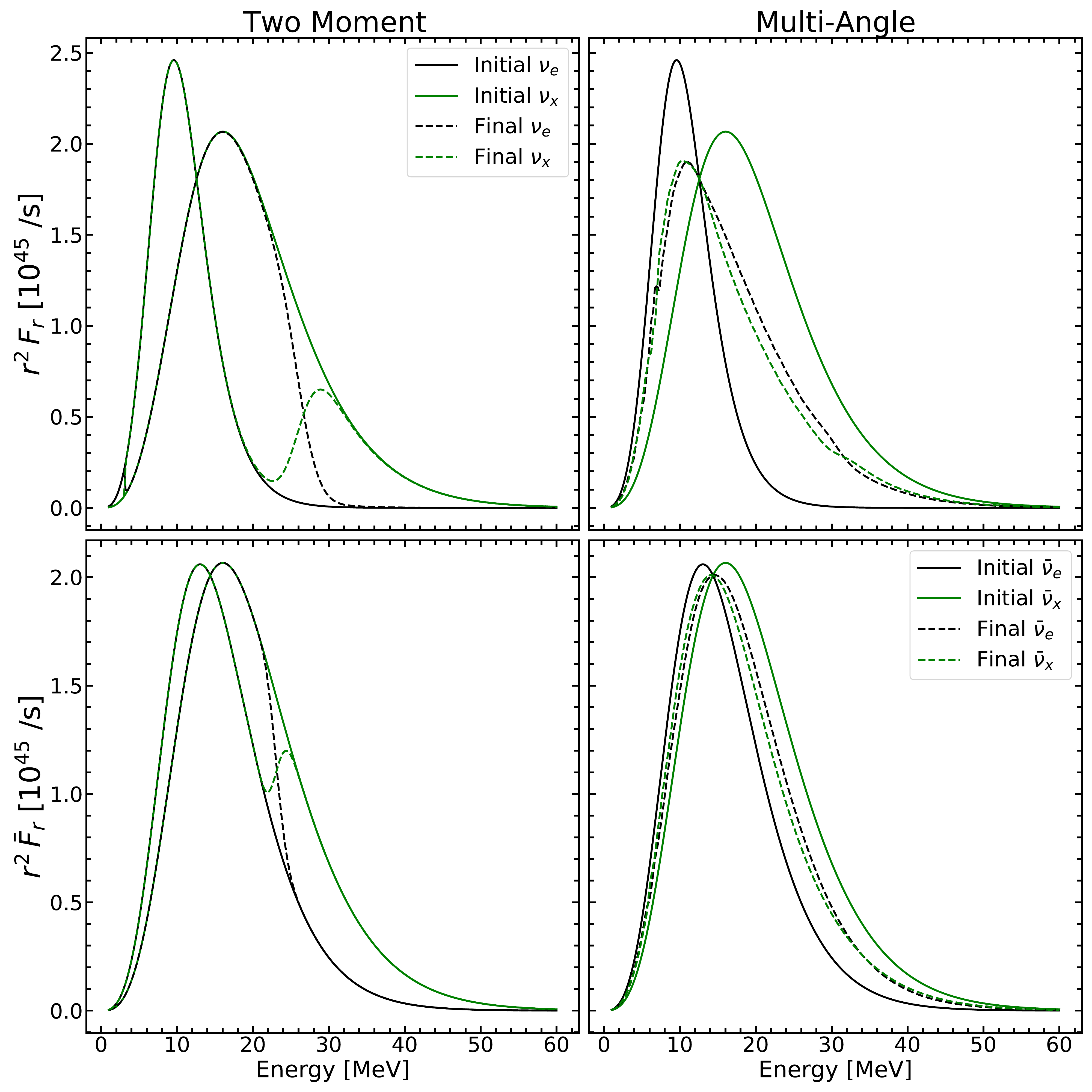}
\caption{Initial and final spectra versus energy for neutrinos from simulations using parameters in Table~\ref{tab:setup_tableIII} and Section~\ref{sec:collective3}.  Neutrinos are in the top two panels and antineutrinos are in the bottom two.  The left column shows the two-moment simulation and the right column shows the multi-angle simulation. The black lines are the electron flavor and the green lines are the $x$ flavor. The solid lines show the initial spectra and the dashed lines show the final spectra.}
\label{fig:spectrum_tableIII}
\end{figure*}

\subsubsection{A case with multiple spectral splits}
\label{sec:collective3}

\begin{table}[b]
    \centering
    \begin{tabular}{l| c c c c c c c}
        a & $L_a$ [ergs/s] & $\langle q \rangle_a$ [MeV] & $T_a$ [MeV] & $\eta_a$ & ${\beta_a}$ & $F_r/E$ & $P_{rr}/E$ \\
        \hline 
    $\nu_{e}$ & $1.8\times10^{52}$ & 12.0 & 2.1 & 3.9 & 0.0 & 0.933 & 0.872 \\
    $\bar{\nu}_{e}$ & $2.2\times 10^{52}$ & 15.0 & 3.5 & 2.3 & 0.0 & 0.933 & 0.872 \\
    $\nu_{x}$ & $2.7 \times 10^{52}$ & 18.0 & 4.4 & 2.1 & 0.0 & 0.933 & 0.872\\
    $\bar{\nu}_{x}$ & $2.7\times10^{52}$ & 18.0 & 4.4 & 2.1 & 0.0 & 0.933 & 0.872
    \end{tabular}
    \caption{Parameters for the collective oscillation simulation discussed in Section~\ref{sec:collective3} which lead to a case of multiple spectral splits. Listed are the luminosity $L_a$, average energy $\langle q \rangle_a$, temperature $T_a$, chemical potential divided by the temperature $\eta_a$, the angular distribution parameter $\beta_a$, the initial flux factor $F_r/E$ and initial Eddington factor and $P_{rr}/E$.}
    \label{tab:setup_tableIII}
\end{table}

Our final test of the moment code is for a case which leads to multiple spectral splits. The parameters for this test problem as shown in table \ref{tab:setup_tableIII} and are a slight adjustment from those used in the first example in \cite{PhysRevLett.103.051105}. The luminosities and mean energies are such that the number fluxes of all four flavors are almost equal, with the electron antineutrino number flux $\sim 2\%$ smaller than the others. This near-equality of the number fluxes leads to a small self-interaction potential which tends to move the instability close to the proto-neutron star.

Figure~\ref{fig:luminosity_tableIII} shows the evolution of the `ee' element of the neutrino and antineutrino luminosity matrices as a function of the radius. A flavor instability is seen to occur around $r\sim 40\;{\rm km}$ and, once again, both moment calculations with the two different closures and the multi-angle calculation are in good agreement about the location of the instability. For the next $\sim 20\;{\rm km}$ thereafter the three calculations track one another closely but at $r\sim 60\;{\rm km}$ there is a noticeable change in the multi-angle calculation compared to the moment-based approach. Beyond $r\sim 60\;{\rm km}$ the multi-angle results exhibit rapid oscillations which decrease in amplitude so that by $r\sim 200\;{\rm km}$ they are no longer observable. The moment-based approaches change noticeably at $r\sim 100\;{\rm km}$ and while the oscillations in the elements of the luminosity matrix also decrease with distance, it is over a much longer scale. In contrast to the two previous self-interaction test problems, there are substantial difference of the luminosities at large radii of the different approaches.

We have explored why the multi-angle and moment based results differ in this test problem by focusing upon the coherence of the two contributions to the self-interaction. Figure~\ref{fig:T3phases} shows the phase angles $\xi_E$ and $\xi_F$ of the off-diagonal elements of the two contributions to the self-interaction, $H_E$ and $H_F$, respectively (i.e.\ $\xi_{E} = \arg(H_E^{e\mu})$, $\xi_{E} = \arg(H_F^{e\mu})$ ) and the difference  between them for the moment-based approach using the two-moment closure, and from the multi-angle calculation. The evolution of these phase angles separately with radius indicate the off-diagonal elements of $H_E$ and $H_F$ rotate rapidly in the Argand plane, but even so, for $r \lesssim 60\;{\rm km}$ the difference between the phase angles is miniscule and the two terms are coherent. Beyond $r \sim 60\;{\rm km}$ the evolution of the phase angles for the multi-angle calculation changes noticeably. At first glance the change in behavior of each phase angle appears to be similar but closer inspection reveals this not to be the case and the difference between the two phases becomes non-zero indicating that actually the degree of coherence between the two contributions to $H_{SI}$ is weaker. The loss of coherence of these two contributions to the off-diagonal elements of the self-interaction Hamiltonian reduces the amount of flavor transformation in the multi-angle calculations beyond $r \sim 60\;{\rm km}$. Fluctuations in $\xi_F - \xi_E$ are not seen in the moment-based results and thus the two contributions to $H_{SI}$ remain coherent leading to significant flavor transformation after $r\sim 60 \;{\rm km}$. In order to verify this difference between the two approaches we have repeated the calculations using different numbers of energy and angle bins, and even a different code, and found the same results every time. Thus we suspect the loss of coherence in the multi-angle calculation is due to the significant cancellation in the self-interaction Hamiltonian due to the spectral parameters chosen for this test problem. Less cancellation - as in the previous two self-interaction test problems - leads to stronger coherence and better agreement between the moment-based and multi-angle results. 

The differences between the two approaches is also seen in the spectra shown in Figure~\ref{fig:spectrum_tableIII}. The spectra from the moment calculation shows two splits in the neutrinos at $E \sim 4\;{\rm MeV}$ and $E \sim 30\;{\rm MeV}$, and $E \sim 25\;{\rm MeV}$ in the antineutrinos, with complete swaps of the spectra outside the split regions. In contrast, the spectra from the multi-angle calculation are very close to a 50:50 mixture of the initial spectra albeit with observable changes in the exact amount of mixing at the same split energies seen more clearly in the moment-based approach. These results are a close match to those seen in \cite{PhysRevLett.103.051105} where single-angle calculations were seen to produce sharp splits and complete swaps, while the splits in the multi-angle calculations were less sharp and the swaps incomplete.

\section{Discussion \& Conclusions}
\label{sec:conclusions}

We have presented here a method for modeling neutrino flavor mixing using one and two moments. We then investigated how well it captures the mixing effects of analytic predictions and of more exact but more computationally expensive multi-angle calculations by considering several problems in a bulb-model geometry representative of core-collapse supernovae. Overall, the moment-based method reproduces results from a multi-angle method surprisingly well. However, errors can emerge in certain circumstances as a result of the tendency of the moment method to maintain an artificially high level of coherence between the simulated moments when using a scalar closure independent of the level at which we truncate the evolution equation tower. This supports the claim that the largest source of error is the nature of the closure and not the number of evolved moments.

The first problem was the flavor evolution of neutrinos emitted from a neutrino bulb in matter with a density chosen to put the 1 MeV neutrinos on the MSW resonance for the given mixing parameters. We found that the moment calculations overestimated the amplitude of the flavor mixing at a given radial point because they were not able to accurately account for the incoherence of neutrinos which had traveled along different trajectories. As the emission at the neutrinosphere becomes more forward-peaked i.e. a greater proportion of the neutrinos are emitted along trajectories close to the radial direction, the transition probabilities from the multi-angle calculation become more similar with those from the moment calculation. 

The last three test problems were used to study the case of collective flavor transformation due to neutrino self interaction. In the first of these calculations, designed with a reasonably realistic inner boundary condition, the moment and multi-angle results were in good agreement in that the onset of the transformation differed by just a few kilometers. The results from the one-moment and two-moment calculations were almost identical. When we adjusted the parameters so as to produce flavor transformation much closer to the neutrinosphere (i.e., where the flux and Eddington factors are much lower), we again found the multi-angle and moment calculations were in good agreement on where the flavor transformation begins and that, once again, the one-moment and two-moment calculations yield essentially the same radial evolution of the luminosities. We then demonstrated that the similarity of the two moment-based approaches is due to a sustained synchronization between the flux and pressure moments which effectively collapses the tower of moment equations to that for a single moment. This coherence between the flux and pressure moments appears to emerge naturally in steady state self-interaction situations. As a final attempt to elicit different outcomes from different methods, we simulate conditions constructed to produce multiple spectral swaps. Once again, the moment method correctly predicted the onset of instability and for some range thereafter, the two approaches were in strong agreement. But eventually the two approaches diverged due to the artificially high levels of coherence in the moment approach compared to the multi-angle and this, in this case, the difference resulted in much larger errors.

Overall, the various tests we undertook indicate that a moment-based approach does well at capturing the overall neutrino flavor transformation seen in the more computationally expensive, multi-angle, calculation. The onset of flavor mixing is predicted with an accuracy of a few kilometers and the spectra are similar. However, we do find differences between the two approaches 
indicating further studies are required before we can completely enjoy the benefits of the moment-based approach. 
Our second self-interaction test problem revealed that the moment-based approaches can yield, at least temporarily, unphysical solutions, and we also saw in our third self-interaction test case how over-estimation of the coherence can lead to different final spectra. 
The differences between the multi-angle and moment-based results are entirely due to the closure. 
The closures we have used in this paper are scalar relations, but for quantum moments in general, one would expect phase differences between the off-diagonal elements of the two moments used in the closure. A scalar closure like those used here cannot generate such a phase difference. To permit the phase differences to appear, one requires a more general, quantum closure. 
We shall explore quantum closures in future work. 

In summary, our results indicate that moment-based schemes are an inexpensive approach to neutrino transport that are able to capture most of the flavor mixing phenomenology seen in more-exact, but more expensive, calculations. We caution that any enthusiasm for a moment-based approach to neutrino oscillations must be tempered by remembering that that any approach which truncates the tower of moment equations cannot converge to those from the full quantum kinetic equations and thus moment-based approaches will continue to require verification against those derived from less approximate methods. Should further, more demanding comparisons of moments and less approximate methods, (especially comparisons which do not assume a steady state) reveal that moment-based methods perform well in those situations, too, we would proffer moment methods are a promising and viable avenue for including neutrino transformation in hydrodynamical simulations.

\acknowledgements{The authors would like to thank S.~Couch and E.~P.~O'Connor for their advice and support through this project. MLW was supported by a NSF Astronomy and Astrophysics Postdoctoral Fellowship under award AST-1801844. TC, MM, JPK, GCM, EG, and CF are supported by the DOE Office of Nuclear Physics award DE-FG02-02ER41216. E. G. acknowledges support by the National Science Foundation grant  No. PHY-1430152 (JINA Center for the Evolution of the Elements). This work was partially enabled by the National Science Foundation under grants No. PHY-1630782 (N3AS hub)  and No. PHY-2020275 (N3AS PFC). SR was supported by a NSF Astronomy and Astrophysics Postdoctoral Fellowship under award AST-2001760. The calculations presented in this paper were undertaken with the support of Michigan State University through computational resources provided by the Institute for Cyber-Enabled Research, and on the \emph{Payne} machine at NCSU which is supported in part by the Research Corporation for Science Advancement. The authors would like to acknowledge the use of the following software: Matplotlib \cite{hunter:2007}, Numpy \cite{vanderwalt:2011}, and SciPy \cite{scipy}.}

\appendix

\begin{figure}[b]
    \centering
    \includegraphics[width=\linewidth]{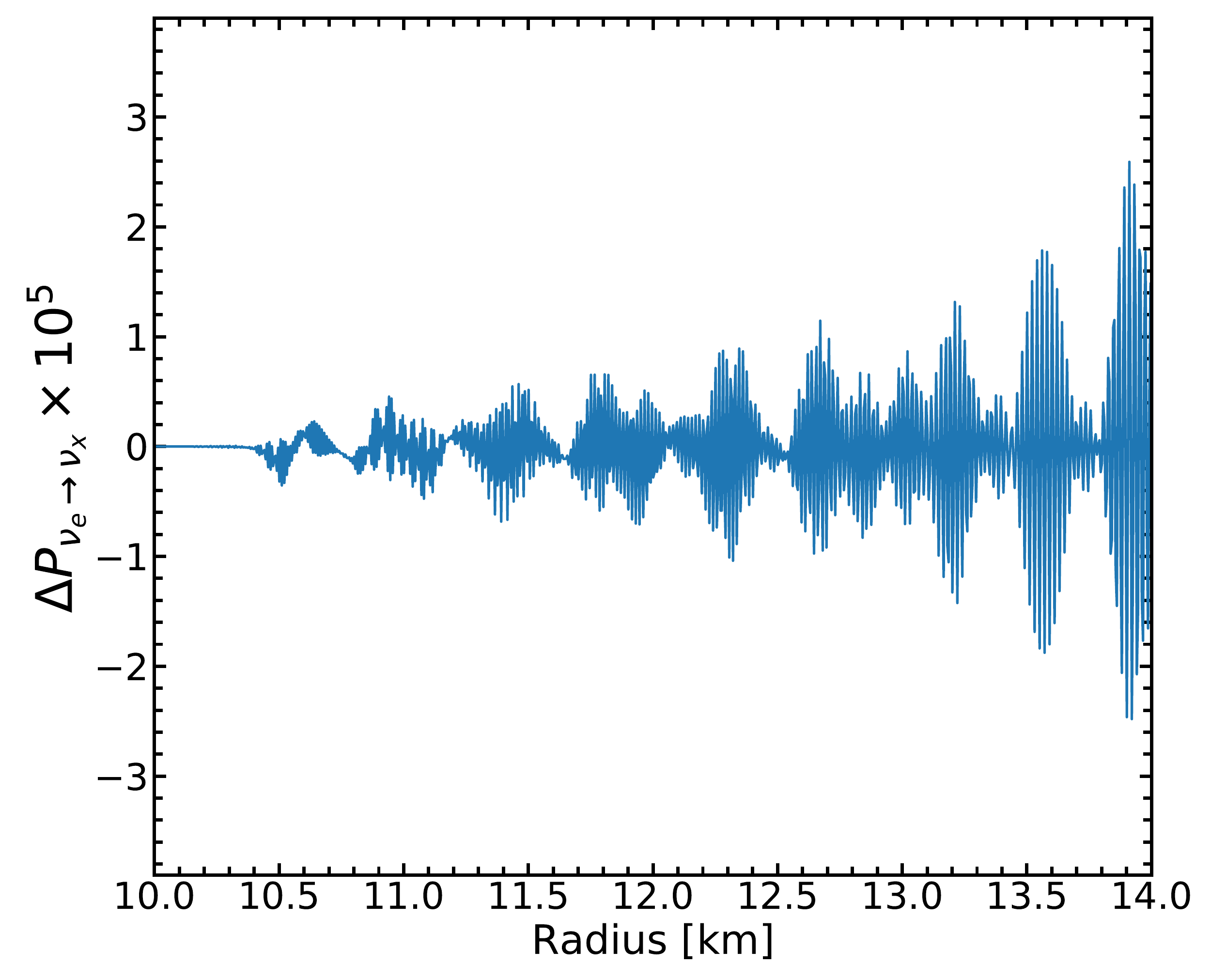}
    \caption{The difference in the transition probability as a function of radius for the 15 MeV neutrinos from a two-moment calculation with an energy resolution of $50\;{\rm keV}$ compared to the fiducial energy resolution of $100\;{\rm keV}$. The calculation uses the parameters shown in Table \ref{tab:setup_tableII}. }
    \label{fig:DeltaP15MeV}
\end{figure}

\section{SQA~\label{sec:SQA}}

For the multi-angle calculations we use SQA, which is an implementation of the Bulb Model described in \citet{duan_coherent_2006}. The momenta of the neutrinos and antineutrinos at the neutrinosphere are discretized into $N_E$ energy bins and $N_A$ angle bins. The energy resolution is uniform, but for the direction we adopt uniform resolution in the quantity $u=\sin^2 \theta_{R}$ where $\theta_R$ is the angle at which the neutrino was emitted at the neutrinosphere relative to the radial direction. This distribution gives greater weight to the angles emitted at large angles $\theta_{R}$, since these rays have the largest dispersion in path lengths from the neutrinosphere to a given radial point. We introduce an evolution matrix $S$ that relates the initial density matrix for neutrinos moving with each momentum $\vec{p}_0$ at the neutrinosphere to the density matrix for neutrinos moving at some distance $\lambda$ along the same trajectory with momentum $\vec{p}(\lambda)$. That is, 
\begin{equation}
    \rho(\lambda,\vec{p}_0) = S(\lambda,\vec{p}_0)\, \rho(R_{\nu},\vec{p}_0) \, S^{\dagger}(\lambda,\vec{p}_0).
\end{equation}
A matrix $\bar{S}(\lambda,\vec{p}_0)$ plays the same role for antineutrinos. The matrix $S(\lambda,\vec{p}_0)$ evolves according to the Schr\"odinger equation
\begin{equation}
i\frac{dS}{d\lambda} = H\left(\lambda, \vec{p} \right) S.
\end{equation}
For the antineutrinos, the matrix $\bar{S}$ similarly evolves according to the Hamiltonian $\bar{H}$. 
As in the moment method described in Section~\ref{sec:flavor_moment_scheme}, the Hamilton is composed of three parts: vacuum $H_V$, matter $H_M$, and self-interaction $H_\mathrm{SI}$. The vacuum and matter terms were given previously in Equations~\ref{eq:H_0} and \ref{eq:H_e}, and the self interaction term is 
\begin{equation}
H_\mathrm{SI}\left( r,{\vec{p}} \right) = \sqrt{2}\,G_{F}\int {\left( {1 - {\bf{\hat p}} \cdot {\bf{\hat q}}} \right)\left[ {\rho(r,\vec{q}) - \bar\rho^ * (r,\vec{q})} \right] q^2 dq\,d\Omega}.
\end{equation}
The density matrices are evolved to discrete radial points $r(\lambda)$ common among all trajectories so the Hamiltonian can be straightforwardly integrated.

To ensure unitarity, the matrices $S$ and $\bar{S}$ are parameterized by four real variables: three are the angles defining a unit vector in a four-dimensional Euclidian flavor space, and the fourth is the phase of the determinant. The differential equations for all of the parameters are solved simultaneously with an explicit Runge-Kutta integrator that uses an adaptive step size and the Cash-Karp parameter set. Several sources of numerical errors in multi-angle codes have been identified over the years \cite{Duan:2006an,2008CS&D....1a5007D,2012PhRvD..86l5020S,2015PhyS...90h8008K}. The numerical accuracy of the SQA calculations presented in this paper is estimated to be less than 1\% based on the lack of visible differences between the results using different numbers of energy or angle bins, and initial radii. 
A related code, IsotropicSQA, was used for solving the neutrino quantum kinetic equations in isotropic and homogeneous conditions \cite{2019PhRvD..99l3014R} is available at \cite{richers_neutrino_2019}.

\section{Numerical Errors in the Moment-Based Calculations~\label{sec:NumericalErrors}}

During our examination of the results from the moment-based code using the two-moment closure, we observed that the luminosites became slightly and temporarily greater than their initial values, as shown in the inset of Figure~\ref{fig:luminosity_tableII}. The same ratio from the code using the one-moment closure scheme did not contain this feature. We have explored the reason for the appearance of these unphysical results from the two-moment scheme and attempted several strategies to remove them. In this appendix we describe those efforts. 

The greater-than-initial luminosities are present only when the self-interaction contribution to the Hamiltonian is included, and they occur for both neutrinos and antineutrinos. We have no reason to suspect that the errors are numerical in origin generated from the ODE integrator we used, since decreasing the error tolerance for the integrator does not improve the solution. We also changed the algorithm from an explicit second-order `midpoint' Runge-Kutta integrator to an adaptive step size routine using the Cash-Karp parameter set and a fractional error tolerance per step of $10^{-10}$, and found the negative factors persisted. We also tried changing the parameterization of the moment matrices by writing the diagonal elements as $M^{(ee)} = Tr(M)\,\cos^2\theta_M$ and $M^{(xx)} = Tr(M)\,\sin^2\theta_M$ where $Tr(M)$ is the trace and $\theta_M$ an angle, in an effort to prohibit the unphysical behavior of the diagonal elements. However, when we ran our code using the two-moment closure to evolve these new parameters, the code terminated prematurely when the step size shrank to zero at the point where the original version crossed into the unphysical solution space. 

\begin{figure}[t]
    \centering
    \includegraphics[width=\linewidth]{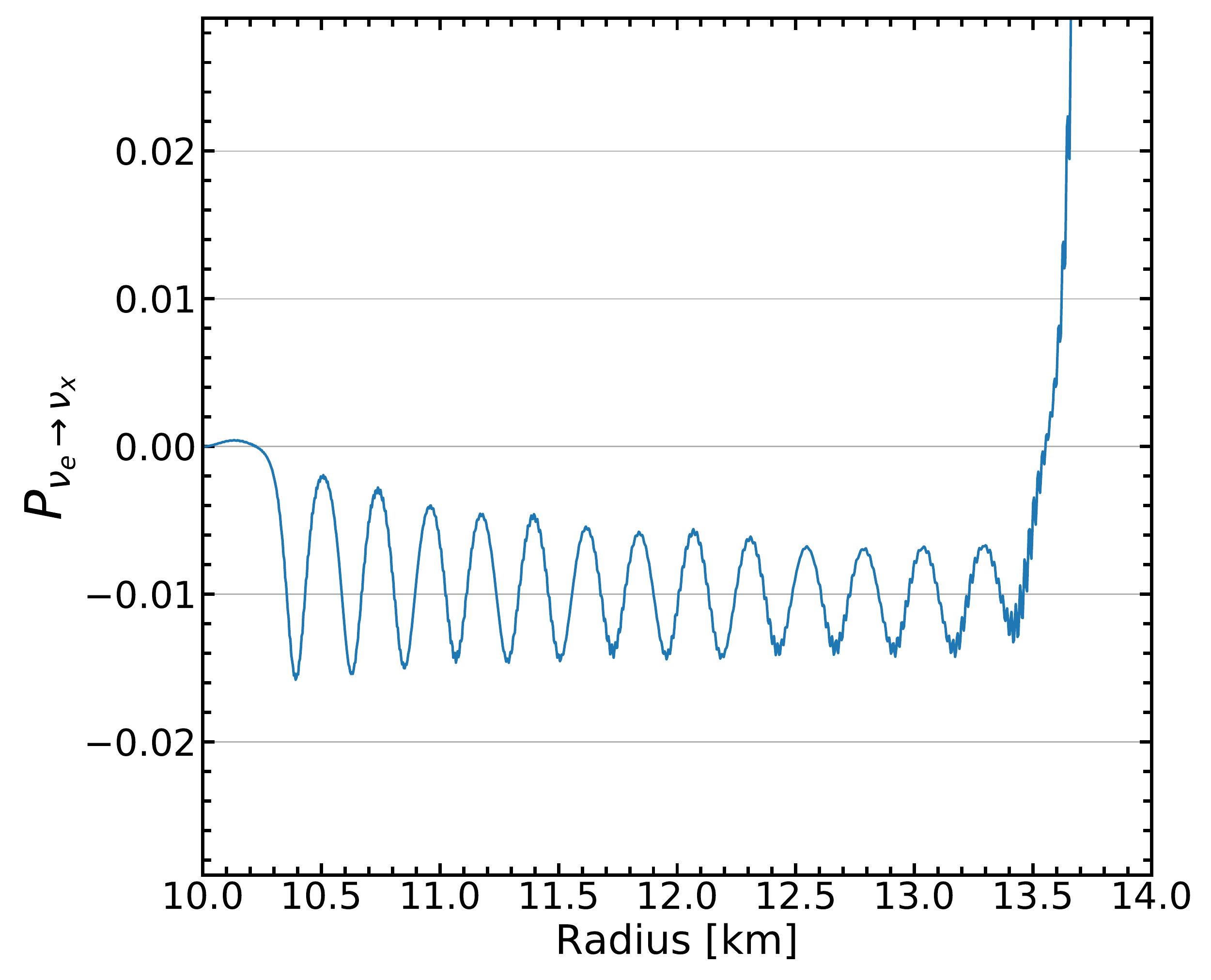}
    \caption{The transition probability as a function of radius as computed from a single-energy calculation using the parameters shown in Table \ref{tab:setup_tableII}. The energy of the neutrinos and antineutrino is chosen to be $5\;{\rm MeV}$ in order to produce an instability in approximately the same location seen in the multi-energy calculation.}
    \label{fig:singleenergy5MeV}
\end{figure}

The greater-than-initial luminosities are not dependent upon the number of energy bins used in the calculation. While the results in Section~\ref{sec:collective2} use and energy resolution of $100\;{\rm keV}$, we repeated the calculation using a energy resolution of $50\;{\rm keV}$. The evolution of the transition probability (the transition probability is defined in Eq. \ref{eq:Ptrans}) for the $15\;{\rm MeV}$ neutrinos from the two calculations is shown in Figure \ref{fig:DeltaP15MeV}. We observe changes in the transition probability that are of order 0.001\% in the interval $10\;{\rm km}\leq r \leq 14\;{\rm km}$ where the greater-than-initial luminosities appear. We also ran a single energy calculation - for which energy resolution is not a factor - using an energy of $5\;{\rm MeV}$. The choice of $5\;{\rm MeV}$ produces an onset of flavor transformation at $r\approx13.5\,\mathrm{km}$, similar to that seen in the multi-energy calculation. The results from this single-energy calculation are shown in Figure (\ref{fig:singleenergy5MeV}) and we observe negative transition probabilities. Thus we do not believe numerical error from the energy resolution to be the origin of the negative factors.

We suspect the origin of the errors to be a combination of using moments, and the closure. In general moments do not evolve according to a unitary operator because a moment is an integrated quantity which aggregates information about the neutrinos traveling along separate trajectories. This is reflected in the evolution equations themselves. The last term on the right hand side of the moment evolution equations (Equations~\ref{eq:energy}-\ref{eq:fluxbar}) unitarily evolves the moment's flavor structure and cannot lead to negative probabilities except for numerical error. The geometrical term (the last term on the left hand side of the equation) leads to non-unitary evolution but this evolution is simply a scaling and cannot lead to the problems we see. It is the first commutator on the right hand side which is the source of physical non-unitary evolution of the moments and depends on the choice of closure. Thus we suspect that the closure we use, while geometrically justified and always realizable, may not be sufficient to enforce physical evolution of the moment equations. An analysis of various options for closures which enforce that the solution always remain physical is beyond the scope of this paper.

\bibliography{paper.bib,SherwoodZotero.bib}

\end{document}